\documentclass[12pt,tightenlines,eqsecnum,floats,aps,superscriptaddress, nofootinbib,prd,showpacs]{revtex4}
\usepackage[T1]{fontenc}
\usepackage[utf8]{inputenc}
\usepackage{lmodern}
\usepackage{amsmath,amssymb,amsfonts}
\usepackage{hyperref}
\usepackage[english]{babel}
\usepackage{graphicx}
\usepackage{color}
\def\be{\nopagebreak[3]\begin{equation}}
\def\ee{\end{equation}}
\def\ba{\nopagebreak[3]\begin{eqnarray}}
\def\ea{\end{eqnarray}}
\def\bas{\nopagebreak[3]\begin{eqnarray*}}
\def\eas{\end{eqnarray*}}

\def\d{{\rm d}}
\def\a{\alpha}

\def\F{{\cal F}}

\def\R{\mathbb{R}}

\def\S{\Sigma}
\def\l{\lambda}

\def\g{\gamma}
\def\R{\mathbb{R}}

\def\ed{\d \!\!\!\! \d\, }

\newcommand{\f}{\frac}

\usepackage{enumerate}

\usepackage{colordvi}
\newcounter{mnotecount}[section]

\def\de{\delta}
\def\o{\omega}
\def\b{\beta}
\def\O{\Omega}
\def\D{\Delta}

\def\ula{\underleftarrow}
\def\heq{\hat{=}}
\def\bm{\bar{m}}
\def\eps{{}^2\!\epsilon}
\def\w{\wedge}
\def\co{8\pi G}

\def\ve{\varepsilon}
\def\Lb{\mathbf{L}}

\def\f{\frac}

\def\La{\pounds}
\def\a{\alpha}

\def\k{\kappa}

\newcommand{\re}{\mathbb{R}}

\def\Po{\mathrm{Po}}
\def\Eu{\mathrm{E}}
\def\ny{\mathrm{NY}}
\def\hny{\mathrm{HNY}}

\begin{document}

\title{Actions, topological terms and boundaries \\
 in first order gravity: A review}
\author{Alejandro Corichi}\email{corichi@matmor.unam.mx}
\affiliation{Centro de Ciencias Matem\'aticas, Universidad Nacional Aut\'onoma de
M\'exico, UNAM-Campus Morelia, A. Postal 61-3, Morelia, Michoac\'an 58090,
Mexico}
\affiliation{Center for Fundamental Theory, Institute for Gravitation and the Cosmos,
Pennsylvania State University, University Park
PA 16802, USA}
\author{Irais Rubalcava-Garc\'\i a}
\email{irais@fcfm.buap.mx}
\affiliation{Facultad de Ciencias F\'{\i}sico Matem\'aticas, Universidad Aut\'onoma de Puebla, 
A. Postal 1152, 72001 Puebla, Puebla, Mexico}
\author{Tatjana Vuka\v sinac}
\email{tatjana@umich.mx}
\affiliation{Facultad de Ingenier\'\i a Civil, Universidad Michoacana de San Nicol\'as de
Hidalgo,\\ Morelia, Michoac\'an 58000, Mexico}
\begin{abstract}
In this review we consider first order gravity in four dimensions. In particular, we focus our attention in 
formulations where the fundamental variables are a tetrad $e_a^I$ and a SO(3,1) connection ${\omega_{aI}}^J$. 
We study the most general action principle compatible with diffeomorphism invariance. 
This implies, in particular, considering besides the standard Einstein-Hilbert-Palatini term, other 
terms that either do not change the equations of motion, or are topological in nature. 
Having a well defined action principle sometimes involves the need for additional boundary 
terms, whose detailed form may depend on the particular boundary conditions at hand. In this work, 
we consider spacetimes that include a boundary at infinity, satisfying asymptotically flat boundary 
conditions and/or an internal boundary satisfying isolated horizons boundary conditions. 
We focus on the covariant Hamiltonian formalism where the phase space $\Gamma$ is given by 
solutions to the equations of motion. For each of the possible terms contributing to the action we 
consider the well posedness of the action, its finiteness, the contribution to the symplectic structure,
and the Hamiltonian and Noether charges. For the chosen boundary conditions, 
standard boundary terms warrant a well posed theory. Furthermore, the boundary and topological terms 
do not contribute to the symplectic structure, nor the Hamiltonian conserved charges. 
The Noether conserved charges, on the other hand, do depend on such additional terms.
The aim of this manuscript is to present a comprehensive and self-contained treatment of the subject, 
so the style is somewhat pedagogical. 
Furthermore, along the way we point out and clarify some issues that have not been clearly understood in the literature.
\end{abstract}
\pacs{04.20.Fy, 04.20.Ha, 04.70.Bw}

\maketitle
\tableofcontents


\section{Introduction}

With the advent of the general theory of relativity it became increasingly clear that
realistic physical theories ought to be formulated, in their Lagrangian description, as 
diffeomorphism invariant theories. This means that one can perform
generic diffeomorphism on the underlying spacetime manifold and the theory remains invariant. 
In most instances, diffeomorphism invariance is achieved by formulating the theory as an action principle 
where the Lagrangian density is defined without 
the use of background structures; it is only the dynamical fields that appear 
in the action. In this manner one incorporates the `stage', the gravitational field, as one of the
dynamical fields that one can describe. The fact that one can write a term that captures the dynamics of 
the gravitational field is noteworthy. An interesting endeavour is to explore the freedom available in 
the definition of an action principle for general relativity. To review these developments
is the main task that we undertake in this manuscript. 
We will restrict ourselves to general relativity as the theory describing the gravitational interaction, and shall 
not consider generalizations such as scalar-tensor theories nor massive gravity in our analysis.

The first issue that one should address is that of having a well posed variational principle. This
particularly  `tame' requirement seems, however, to be sometimes overlooked in the literature. 
It is natural to ask why one needs to have a well posed action principle when, at the end of the day, 
we already `know' what the field equations are. While this might be the case, one should keep in 
mind that the classical theory is only a (very useful indeed!) approximation to a deeper underlying theory
that must be quantum in nature. If one takes the viewpoint that at the deepest level, 
any physical system is quantum mechanical and can be
defined by some path integral, in order for this to be well defined, we need to 
write a meaningful, finite, action. That is, one should be able to define an 
action for the whole space of histories, and not only for  classical solutions. 
This simple observation becomes 
particularly important when the physical situation under study involves a spacetime region with 
boundaries. In this case, one must 
be  careful to extend the formalism in order to incorporate boundary terms.

Another equally important issue in the definition of any physical theory is the choice of fundamental 
variables, and even more
when gauge symmetries are present. This issue is particularly important. For, even when the space of 
solutions might coincide for two formulations, 
the corresponding actions will generically be different and that will have an effect in the path integral formulation of the theory. In the case of general relativity, the original and better known formulation, as conceived by Einstein, is written in terms of a metric 
tensor $g_{ab}$ satisfying second order equations \cite{einstein}. As is well known, other choices of variables might yield alternative descriptions. In this review, we shall explore one of those possibilities. In particular, if one wants to couple fermions to the gravitational field (a very reasonable request), then the second order formalism does not suffice. One needs to consider instead co-tetrads $e_a^I$  that can be regarded as a ``square root of the metric'': $g_{ab}=e_a^Ie_b^J\eta_{IJ}$. As a byproduct, this choice also allows to cast the theory as a {\it local} gauge theory under the Lorentz group.
It has been known for a long time that one can either obtain Einstein equations of motion by means of the Einstein Hilbert action or in terms of the so called Palatini action, a first order action 
in terms of tetrads $e_a^I$ and a connection ${\omega_{aJ}}^I$ valued on the Lie algebra of $SO(3,1)$ (see. e.g. \cite{romano} and \cite{peldan})\footnote{One should recall that the original Palatini action was written in terms of the metric $g_{ab}$ and an affine connection ${\Gamma^a}_{bc}$ \cite{palatini,adm}. The action we are considering here, in the so called ``vielbein'' formalism, was developed in \cite{utiyama,kibble,sciama} and in \cite{di} in the canonical formulation.}.
Furthermore, one can generalize this action by adding a term, the so-called Holst term, that yields the same equations of motion. 
This  `Holst action' is the starting point for loop quantum gravity and some spin foam models, given that one can describe the theory, in its canonical decomposition in terms of a real $SU(2)$ connection  (see. e.g. \cite{holst}  and \cite{barros}). 
 
Within the same ``vielbein'' scheme one can consider the most general diffeomorphism invariant first order action that classically describes general relativity. It can be written as the Palatini action (including the Holst term) plus topological contributions, namely, the Pontryagin, Euler and Nieh-Yan terms (see for instance \cite{topo} for early references). 

One of the main themes that we want to explore in this manuscript is the case when the spacetime region under consideration possesses boundaries. The main consequence of this choice is that one might have 
to add extra terms (apart from the topological terms that can also be seen as boundary terms) to the action principle so that it becomes well defined\footnote{One should clarify the use of  `topological term'. In this manuscript, a term is topological if it can be written as a total derivative. This  implies that it does not contribute to the equations of motion. There are other possible terms that do not contribute to the equations of motion but that can not be written as a total derivative (such as the so called Holst term). 
Thus, according to our convention, the Holst term is not topological even when by itself it possesses no local degrees of freedom \cite{merced-holst}.}.

With this assumption in mind, the most general first order action for gravity can be written as,
\begin{equation}\label{1}
S[e, \omega] = S_{\rm Palatini} + S_{\rm Holst Term} + S_{\rm Pontryagin} + S_{\rm Euler} + S_{\rm Nieh-Yan} + S_{\rm Boundary}.
\end{equation}
It is important to emphasize that in the  textbook treatment of Hamiltonian systems one usually considers compact spaces without boundary, so there are no boundary terms  coming from the integration by parts in the variational principle. If
one is interested in spacetimes with boundaries,  these boundary terms need to be considered and  analyzed with due care. One then requests
the action principle to be well posed, i.e. one requires the action to be {\em differentiable 
and finite} under the appropriate boundary conditions, and under the most general variations compatible with the boundary conditions. 
Indeed some progress has been done in this direction. 
Under appropriate boundary conditions\footnote{See e.g. \cite{aes}, \cite{afk}, \cite{apv} and references therein for the asymptotically flat, isolated horizons and asymptotically AdS spacetimes respectively.}, 
the Palatini action plus a boundary term provides a well posed action principle, that is, it is differentiable and finite. An explicitly gauge invariant boundary term, useful for finite boundaries, was put forward in \cite{bn}. Furthermore, in \cite{cwe} the analysis for 
asymptotically flat boundary conditions was extended to include the Holst term. Isolated horizons boundary conditions were studied in \cite{cg} and \cite{crv1}.

The covariant Hamiltonian formalism (see, e.g. \cite{abr}, \cite{witten} and \cite{wald-lee}) seems to be particularly well suited for
exploring relevant properties of the theories defined by an action principle. In this formalism, one can introduce standard Hamiltonian structures such as a phase space, symplectic structure, canonical transformations, 
 without the need of a $3+1$ decomposition of the theory. 
All the physical quantities  are defined in a {\it covariant} way. One of the most attractive feature of this formalism is that one can find all 
these structures
in a unique fashion given the action principle. Even more, conserved quantities can be found in a `canonical' way. 
On  the one hand one can derive Hamiltonian generators of 
canonical transformations and, on the other hand, Noetherian conserved quantities associated to symmetries. 
Of the most importance is to understand the precise relation between these two sets of quantities. We shall review this relation here.

The study of field theories with boundaries within the Hamiltonian approach is certainly not new in the 
literature. However, most of these studies focus on the 3+1 formalism, where a decomposition is involved 
and constraints are present. One recent study of the role of boundaries in linear field theories, both in the 
canonical and covariant Hamiltonian frameworks, is given in \cite{barbero1}.
General relativity in the second order formulation has indeed been studied in the context of a $3+1$ 
decomposition with asymptotically flat boundary conditions. The first proposal of a boundary term to 
supplement the Einstein Hilbert action came from Gibbons and Hawking \cite{GH} and independently by 
Brown and York  \cite{York}. A summary of such approaches was given by Hawking and Horowitz \cite{HH} 
(See also \cite{York2}). This approach, however, suffers from a lack of generality given that it depends on 
the ability to embed a three dimensional hypersurface in spacetime. A procedure to deal with asymptotically 
flat configurations and that overcomes this limitation was recently put forward by Mann and Marolf 
\cite{MM}. A detailed study of
the 3+1 decomposition of a first order gravity action with asymptotically flat boundary conditions was only 
recently completed \cite{C-JD}.

The purpose of this manuscript is to review all the issues we have mentioned in a systematic way, within the first order formalism. More concretely, we have three main goals. The first one is to explore the well-posedness of the action principle with boundary terms. For that we study two sets of boundary conditions that are physically interesting; as outer boundary we consider 
configurations that are asymptotically flat. For an inner boundary, we consider those histories that satisfy isolated horizon boundary conditions. The second objective is to explore the 
most basic structures in the covariant phase space formulation. More precisely, we study the existence of the symplectic structure as a finite quantity and its dependence on 
the various topological and boundary terms. Finally, the last goal of this manuscript is to revisit the different conserved quantities that can be defined. Concretely, we consider Hamiltonian 
conserved charges both at infinity and at the horizon. Finally, we compare them with the associated Noetherian conserved current and charges. 
In both cases we review in detail how these quantities depend on the existence of the boundary terms that make the action well defined. As it turns out, while the 
Hamiltonian charges are insensitive to those quantities, the Noether charges {\it do} depend on the form of the boundary terms added. While the main objective of this manuscript is to review material that
has appeared elsewhere, we include some new results and clarifications of several issues. Since, to the best of our knowledge, there is no reference where all these results have
been put in a coherent and systematic fashion, the final goal of this contribution is to fill this gap and  present the subject in a pedagogical and self-contained manner.   

The structure of the manuscript is as follows:
In Section~\ref{sec:2} we review what it means for an action principle to be well posed, which is when it is finite and differentiable.
In Section~\ref{sec:3} we use some results discussed in the previous section, to review the covariant Hamiltonian formalism taking enough care in the cases when the spacetime has 
boundaries. We begin by defining the covariant phase space and its relation with the canonical phase space. Then we introduce the symplectic structure with its 
ambiguities and its dependence on boundary terms in the action. Finally we define the symplectic
current, symplectic structure and, in the last part, recall the definition of the Hamiltonian 
and Noether charges.
In Section~\ref{sec:4} we use the covariant Hamiltonian formalism to study the action introduced in Eq.~(\ref{1}). 
We find the generic boundary terms that appear when we vary the different components of the action.
In Section~\ref{sec:5} we consider particular choices of boundary conditions in the action principle. More precisely, we study spacetimes with boundaries: Asymptotically flatness at 
the outer boundary, and an isolated horizon as an internal one.
In Section~\ref{sec:6} we study symmetries and their generators for both sets of boundary conditions. In particular we first compute the Hamiltonian conserved charges, and in the 
second part, the corresponding Noetherian quantities are found. We comment on the difference between  them.
We summarize and provide some discussion in the final Section~\ref{sec:7}.


\section{Action principle}
\label{sec:2}

In this section we review the action principle that plays a fundamental role in the formulation of physical 
theories. In order to do that we need to be precise about what it means to have a well posed variational 
principle. In particular, there are two aspects to it. The first one is to define the action by itself. 
This is done in the first part of this section. In the second part, we introduce the variational principle 
that states that physical configurations will be those that make the action stationary. In particular, we 
entertain the possibility that the spacetime region under consideration has non-trivial boundaries and that 
the allowed field configurations can vary on these boundaries. These new features require an 
extension of the standard, textbook,  treatment.

\subsection{The action}
\label{sec:2.1}

In particle mechanics the dynamics is specified by some action, which is a {\it function} of the trajectories 
of the particle. In turn, the action $S$ is the time integral of the Lagrangian function $L$ that generically 
depends on the coordinates and velocities of the particles. In field theory the dynamical variables, the fields, 
are geometrical objects defined on spacetime; now the Lagrangian has as domain this function space. 
In both cases, this type of objects are known as functionals. In order to properly define the action we will 
review what is a functional and some of its relevant properties.

A {\it functional} is a map from a \emph{normed space} (a vector space with a non-negative real-valued 
norm\footnote{We need the concept of the norm of a functional to have a notion of closedness and therefore continuity 
and differentiability, for more details see e.g. chapter 23 of \cite{km}.}) 
into its underlying field, which in physical applications is the field of the real numbers. This vector space 
is normally a functional space, which is why sometimes a functional is considered as a function of a function.

A special class of 
functionals are the definite integrals that define an action by an expression of the form,
\begin{equation}\label{DefAction}
S[\phi] = \int_{\mathcal{M}} \mathcal{L} (\phi^{\alpha}, \nabla\phi^\alpha,...,\nabla^n\phi^{\alpha})\,\d^4 V,
\end{equation}
where $\phi^{\alpha}(x)$ are fields on spacetime, $\widetilde{\mathcal{M}}$, ${\mathcal{M}}\subseteq\widetilde{\mathcal{M}}$
is a spacetime region, $\alpha$ is an abstract label for 
spacetime and internal indices\footnote{Throughout the manuscript we shall use Penrose's abstract index notation.}, $\nabla\phi^\alpha$ their first derivatives, and $\nabla^n\phi^\alpha$ 
their $n^{\rm th}$ derivatives, and $\d^4 V$ a volume element on spacetime.
This integral $S[\phi]$ maps a  field history $\phi^\alpha(x)$ into a real number if the Lagrangian 
density $\mathcal{L}$ is real-valued.

Prior to checking the well posedness of this action, we will review what it means for an action to be 
finite and differentiable. We say that an action 
is \emph{finite} iff the integral that defines it 
is convergent or has a finite value when evaluated in {\em histories} compatible with the boundary conditions.

\subsection{Differentiability and the variational principle}
\label{sec:2.2}

As the minimum action principle states, the classical trajectories followed by the system are 
those for which the action is a stationary point. This means that, to first order, the variations 
of the action vanish. As is well known, the origin of this emphasis on extremal histories comes from
the path integral formalism where one can show that trajectories that extremise the action contribute the most to the path integral.
First, let us consider some definitions:

Let $\F$ be a normed space of functions. A functional $F : \F\to \R$ is called differentiable if we 
can write the finite change of the action, under the variation $\phi\to\phi+\de\phi$, as
\be
F [\phi +\de\phi ]-F [\phi ]=\de F + R\, ,
\ee
where $\de\phi\in\F$ (we are assuming here that vectors $\de \phi$ belong to the space $\F$, so it 
is a linear space). The quantity $\de F [\phi ,\de\phi ]$ 
depends linearly on $\de\phi$, and $R[\phi ,\de\phi]
={\mathcal O}((\de\phi)^2)$.
The linear part of the increment, $\de F$, is called the {\em  variation of the funcional}  $F$ (along $\de\phi$).
A stationary point $\bar\phi$ of a differentiable funcional $F [\phi ]$ is a function $\bar\phi$ such that
$\de F[\bar\phi ,\de\phi]=0$ for all $\de\phi$. 

As is standard in theoretical physics, we begin with a basic assumption: The dynamics is specified by an 
action. In most field theories the action depends only on the fundamental fields and their first derivatives. 
Interestingly, this is not the case for the Einstein Hilbert action of general relativity, but it {\em is} 
true for first order formulations of general relativity, which is the case that we shall analyze in the present work.

In general, we can define an action on a spacetime region  $\mathcal{M}$
depending on the fields, $\phi^{\alpha}$ and their first derivatives, $\nabla_{\mu} \phi^{\alpha}$. 
Thus, we have
\begin{equation}\label{action}
S[\phi^{\alpha}] = \int_{\mathcal{M}} \mathcal{L} \left( \phi^{\alpha}, \nabla_{\mu} \phi^{\alpha} \right)\, \d^4 V\, .
\end{equation}
Its variation $\de S$ is the linear part of
\be
\int_{\mathcal{M}}\, \bigl[ \mathcal{L} \left( {\phi '}^\a , \nabla_{\mu} {\phi '}^\a \right)
- \mathcal{L} \left( \phi^{\alpha}, \nabla_{\mu} \phi^{\alpha} \right)\bigr]\, \d^4 V\, ,
\ee
where ${\phi '}^\a =\phi^{\alpha}+\de\phi^{\alpha}$. It follows that
\begin{equation}
\delta S[\phi^{\alpha}] = \int_{\mathcal{M}} \left[ \frac{\partial \mathcal{L} }{\partial \phi^{\alpha}} - 
\nabla_{\mu} \frac{\partial \mathcal{L} }{\partial ( \nabla_{\mu} \phi^{\alpha})} \right] \delta \phi^{\alpha} 
\,\, \d^4 V + \int_{\mathcal{M}} \nabla_{\mu} \left( \frac{\partial \mathcal{L} }{\partial ( \nabla_{\mu} \phi^{\alpha})} \,
\delta \phi^{\alpha} \right) \d^4 V\, ,
\end{equation}
where we have integrated by parts to obtain the second term. 
Let us denote the integrand of the first term as: 
$E_{\alpha} := \frac{\partial \mathcal{L} }{\partial 
\phi^{\alpha}} -
\nabla_{\mu} \left( \frac{\partial \mathcal{L} }{\partial ( \nabla_{\mu} \phi^{\alpha})} \right)$.
Note that the second term on the right hand side is a divergence so we can write it as a boundary term using Stokes' theorem,
\begin{equation}\label{StandardSymplecticPotential}
\int_{\partial \mathcal{M}} \frac{\partial \mathcal{L} }{\partial ( \nabla_{\mu} \phi^{\alpha})}\, \delta \phi^{\alpha} \;
\d S_\mu =: \int_{\partial \mathcal{M}} \theta (\phi^{\alpha}, \nabla_{\mu} \phi^{\alpha}, \delta \phi^{\alpha}) \; \d^3 v\, ,
\end{equation}
where we have introduced the quantity $\theta$ that will be relevant in sections to follow. Note that the 
quantity $\delta S[\phi^{\alpha}]$ can be interpeted as the directional derivative of the funtion(al) $S$ along 
the vector $\de\phi$. Let us introduce the simbol $\ed$ to denote the exterior derivative on the functional 
space ${\cal F}$. Then, we can write $\de S[\phi]=\ed\, S(\de\phi)=\de\phi(S)$, where the last equality 
employs the standard convention of representing the vector field, $\de\phi$, acting on the function $S$.

As we mentioned before, if we want to derive in a consistent way the equations of motion for the system, 
the action must be differentiable. In particular, this means that we need the boundary term 
(\ref{StandardSymplecticPotential}) to be zero. To simply demand that  
$\delta \phi^{\alpha} |_{\partial \mathcal{M}} = 0$, 
as is usually done in introductory textbooks, becomes too restrictive if we want to allow {\em all}  
the variations $\delta\phi^\alpha$ which preserve appropiate boundary conditions and not just variations 
of compact support. Thus, requiring the action to be stationary with respect to all compatible variations 
should yield precisely the classical equations of motion, with the respective boundary term vanishing 
on any allowed variation.

\begin{figure}[h]
\begin{center}
  \includegraphics[width=8cm]{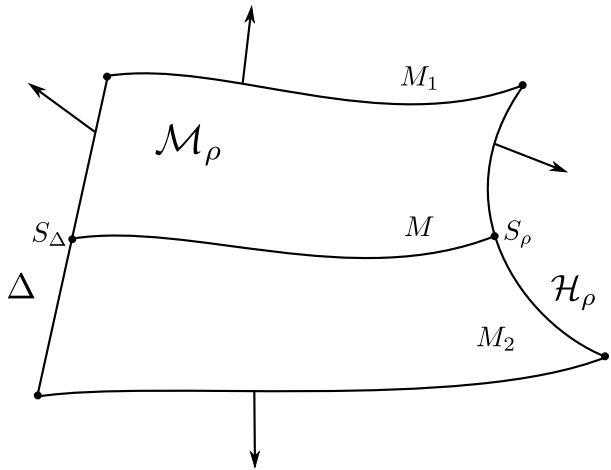}
  \caption{The region ${\mathcal{M}}_\rho$ bounded by two space-like hypersurfaces $M_1$ and $M_2$, a time-like hyperboloid ${\mathcal H_\rho}$ as an
  outer boundary and a hypersurface $\Delta$ as an inner boundary. Corresponding normal vectors are also represented.  }\label{regionM}
  \end{center}
\end{figure}

Let us now consider the case in which the spacetime region ${\mathcal{M}}\subseteq\widetilde{\mathcal{M}}$, where the action is defined, 
has a boundary $\partial{\mathcal M}$. We are interesting in globally hyperbolic asymptotically flat 
spacetimes (so that $\widetilde{\cal M}\approx{\mathbb{R}}\times M$, where $M$ is a space-like non-compact 
hypersurface) possibly with an internal boundary, as would be the case when there is a black hole present.
We can foliate the asymptotic region by time-like hyperboloids ${\cal H}_\rho$, corresponding
to $\rho ={\rm const.}$, and introduce a family of spacetime regions $\{ {\mathcal{M}}_\rho\}_{\rho\in I\subset\mathbb{R}}$,
with a boundary $\partial{\mathcal{M}}_\rho =M_1\cup M_2\cup {\cal H}_\rho\cup \Delta$, where $\Delta$ is an inner
boundary (see Fig.\ref{regionM}). This family satisfy ${\mathcal{M}}_\rho\subset {\mathcal{M}}_{\rho'}$ for $\rho'>\rho$
and ${\mathcal{M}}=\cup_{\rho}{\mathcal{M}}_\rho$.
Then, the integral over ${\cal M}$ in (\ref{action}) is defined as
\begin{equation}
S[\phi^{\alpha}] = \lim_{\rho\to\infty}\int_{\mathcal{M}_\rho} \mathcal{L} \left( \phi^{\alpha}, 
\nabla_{\mu} \phi^{\alpha} \right)\, \d^4 V\, .
\end{equation}

Now, given an action principle and boundary conditions on the fields, a natural question may arise, 
on whether the action principle will be well posed. So far there is no general answer, but there are 
examples where the introduction of a boundary term is needed to make the action principle well defined, 
as we shall show in the examples below. Let us then keep the discussion open and consider  a generic action principle
that we assume to be well defined in a region with boundaries, and with possible contributions to the 
action by boundary terms. Therefore, 
the action of such a well posed variational principle will look like,
\begin{equation}\label{GeneralWellPosedAction}
S[\phi^{\alpha}] = \int_{\mathcal{M}} \mathcal{L} \left( \phi^{\alpha}, \nabla_{\mu} \phi^{\alpha} \right)\, \d^4 V +
\int_{\partial \mathcal{M}} \varphi (\phi^{\alpha}, \nabla_{\mu} \phi^{\alpha}) \,\, \d^3 v\, ,
\end{equation}
where we have considered the possibility that there is contribution to the action coming from the boundary 
$\partial{\cal M}$. Thus, the variation of this {\em extended} action becomes,
\begin{equation}\label{GeneralWellPosedVarPrin}
\delta S[\phi^{\alpha}] = \int_{\mathcal{M}} E_{\alpha}\; \delta \phi^{\alpha} \; \d^4 V +  \int_{\partial \mathcal{M}} 
\theta (\phi^{\alpha}, \nabla_{\mu} \phi^{\alpha}, \delta \phi^{\alpha}) \; \d^3 v +  \int_{\partial \mathcal{M}} 
\delta \varphi (\phi^{\alpha}, \nabla_{\mu} \phi^{\alpha}) \; \d^3 v\, .
\end{equation}
 The action principle will be well posed if the first term is finite and $\varphi (\phi^{\alpha})$ is a 
 boundary term that makes the action well defined under appropriate boundary conditions. That is, when 
 the action is evaluated along {\em histories} that are compatible with the boundary conditions, the 
 numerical value of the integral should be finite, and in the variation (\ref{GeneralWellPosedVarPrin}), 
 the contribution from the boundary terms must vanish. Now, asking 
$\delta S[\phi^{\alpha}]=0$, for arbitrary variations
$\de \phi$ of the fields, implies that the fields must satisfy $$E_\alpha=0\, ,$$
the {\em Euler-Lagrange} equations of motion.

Note that in the ``standard approach'',  one  considers variations, say, of compact support 
such that $\delta \phi^{\alpha} |_{\partial \mathcal{M}} = 0$ (and also the  variations of the derivatives of the fields vanish on the boundary).  
In this case, we can always add a term of the form 
$\nabla_{\mu} \chi ^{\mu}$ to the Lagrangian density,
\begin{equation}
\mathcal{L} \rightarrow \mathcal{L} + \nabla_{\mu} \chi ^{\mu},
\end{equation}
with $\chi$  arbitrary. 
The relevant fact here is that this term will {\em not} modify the equations of motion since the variation 
of the action becomes,
\begin{equation}
\delta S = \delta \int_{\mathcal{M}} \mathcal{L}\; \d^4 V+ 
\delta \int_{ \mathcal{M}} \nabla_{\mu} \chi ^{\mu}\; \d^4 V = \delta \int_{\mathcal{M}} \mathcal{L}\; \d^4 V
+ \int_{\partial \mathcal{M}} \delta \chi ^{\mu}\; \d S_\mu\, ,
\end{equation}
and, since the variations of $\phi^{\alpha}$, as well as their derivatives, vanish on the boundary, 
the second term of the right-hand side always vanishes,
that is,  $\de\chi ^{\mu} |_{\partial \mathcal{M}} = 0$, independently of the detailed form of the resulting boundary term. Therefore, it does not 
matter which boundary term we add to the action; it will not modify the equations of motion. Note that 
within this viewpoint, the action is always assumed to be differentiable from the beginning and the addition of boundary terms does not change this property.

On the contrary, when one considers variational principles of the form (\ref{GeneralWellPosedAction}), 
consistent with arbitrary (compatible) variations in spacetime regions with boundaries, we cannot  just 
add arbitrary total divergences/boundary term to the action, but only those that preserve the action 
principle well-posedness, in the sense mentioned before.
Adding to the action any other term that does not satisfy this condition will spoil the differentiability 
properties of the action and, therefore, one would not obtain the equations of motion in a consistent manner.

This concludes our review of the action principle. Let us now recall how one can get a consistent 
covariant Hamiltonian formulation, once the action principle at hand is well posed.


\section{Covariant Hamiltonian formalism}
\label{sec:3}

In this section we give a self-contained review of the {\em covariant Hamiltonian formalism} (CHF) 
taking special care of the cases where boundaries are present. 

Recall that a theory  has a \emph{well posed initial value formulation}\footnote{We say that a theory possesses 
an initial value formulation if it can be formulated in a way 
that by specifying \emph{appropriate initial data} (maybe restricted to satisfy certain constraints) its 
dynamical evolution is uniquely determined.
For a nice treatment see, e.g., \cite{W} chapter 10.},
if, given initial data there is a unique solution to the equations of motion. In this way there is
an isomorphism ${I}$ between the space of solutions to the equations of motion, $\Gamma$, and the space of 
all admissible initial data, the `canonical phase space' $\Gamma_c$. On this even dimensional space\footnote{Recall that in 
particle mechanics, if we have $n$ particles, we need to specify as initial data their initial 
positions and velocities, so the space of all possible initial data is an even dimensional space. We can easily 
extend this to field theory.}, 
we can construct a nondegenerate, closed 2-form $\Omega_c$, the symplectic form. The pair formed by the phase space 
and the symplectic form constitute a symplectic manifold $(\Gamma_c,\Omega_c)$.

We can bring the symplectic structure to the space of solutions, via the pullback ${I}^*$ of $\Omega_c$ and 
define a corresponding 2-form on $\Gamma$.
Since the Lie derivative of the symplectic structure vanish along the vector field generating time evolution, 
$\Omega_c$ does not depend on the particular choice of the initial instant of time. 
Given that the mapping is independent of the reference Cauchy surface one is using to 
define ${I}$, the space of solutions is equipped with a \emph{natural}
symplectic form, $\Omega$. The space of solutions and its symplectic structure $(\Gamma, \Omega)$ are known
as the \emph{covariant phase space} (CPS) (For early references see \cite{abr,witten}).

Interestingly, most of the field theories of physical relevance posses \emph{gauge} symmetries. This feature of the system has important 
consequences. To begin with, the isomorphism $I$ is not well defined since initial data do not uniquely determine a solution of the
Lagrangian equations of motion. In this case the covariant $\Omega$ is constructed directly from the action principle, as we shall see below.
Furthermore, not all initial data is allowed, and is subject to certain constraints. These two facts
imply that both
symplectic forms $\Omega$ and the restriction $\bar{\Omega}_c$ of the (kinematical) canonical symplectic form to the constraint surface, 
are \emph{degenerate}. One should note that the relation between $\Omega$ and $\bar{\Omega}_c$ is not straightforward and we shall not pursue it here.
When $\Omega$ is degenerate, as shall be the case here, it is called a pre-symplectic 
form. It is only after one gets rid of this degeneracy, by means of an appropriate 
quotient, that one recovers a physical non-degenerate symplectic structure $\hat{\Omega}$. It is at this point that one expects to
recover an isomorphism with the corresponding non-degenerate form $\hat{\Omega}_c$ of the canonical theory (see for instance \cite{wald-lee} for a discussion). 

This section has three parts. In the first one, we define the covariant phase space and its relevant structures, namely the
symplectic potential,  current and structure, starting from the action principle. 
In particular, we analyze the influence of boundary terms 
in the original action and the additional ones that appear in the `variation' of the action.
In the second part, we recall how the symmetries of the underlying spacetime get reflected 
on the covariant Hamiltonian formalism. 
We will pay special attention to the construction of the corresponding conserved 
quantities. These are noteworthy since they are both conserved and play an important role as generators of such symmetries. 
The symmetries that we shall focus on are closely related 
to the issue of diffeomorphism invariance. In the third part we compare  the Hamiltonian 
conserved quantities with the  Noether  charges. We illustrate their  relation and show
that, in contrast to the Hamiltonian charges, these `Noetherian' quantities 
{\em indeed} depend on the existence of boundary terms in the original action.

\subsection{Covariant phase space}
\label{sec:3.1}

In this part we present a review of the covariant phase space and its relevant structures.
From now on we will use the language of differential forms that will prove to be useful and simplify the notation.
However, we need to distinguish between the exterior derivative $\ed$ in 
the infinite dimensional covariant phase space,  and the exterior derivative on the spacetime manifold, denoted by $\d$. 
Note that we shall use $\delta$ or $\delta \phi$ 
to denote tangent vectors on the CPS, to be consistent with the standard notation used in  the literature. Let us now recall some basic constructions on the covariant phase space.

Taking as starting point an action principle, let us first consider the action without any additional boundary term as discussed in the previous section,
\begin{equation}\label{original_action}
S[\phi^{A}] = \int_{\mathcal{M}}  \mathbf{L}\, ,
\end{equation}
we can consider the Lagrangian density, $\mathbf{L}$, as a $4-$form and the fields $\phi^{A}$ as certain $n-$forms (with $n \leq 4$) in the 
$4-$dimensional spacetime manifold. Recall that in the previous section we used $\alpha$ as a generic (abstract) index that could be space-time 
or internal. In the language of forms the spacetime index referring to the nature of the object in space-time will not appear explicitly, so we 
are left only with internal indices that we shall denote with $A,B,\ldots$ to distinguish them from spacetime indices $\mu,\nu,\ldots$. 
Then, the variation of the action can be written as (\ref{GeneralWellPosedVarPrin}) or, equivalently in terms of forms as,
\begin{equation}\label{VarActFormsWithoutBoundary}
\ed S (\delta)  = \delta S = \int_{\mathcal{M}} E_{A} \wedge \delta \phi^{A} + \int_{\mathcal{M}} \d \theta ( \delta \phi^{A}),
\end{equation}
where $E_{A}$ are the Euler-Lagrange equations of motion forms and $\delta \phi^{A}$ is an arbitrary vector on the tangent space.
The 1-form (in CPS) $\theta$ depends on $\phi^A$, $\delta\phi^A$ and their derivatives, even when for
simplicity we do not always write it explicitly. Note that we are using $\delta \phi^{A}$ and $\delta$, to denote \emph{the same} object. As we mentioned in the 
previous section, the second term of the RHS is obtained after integration by parts, and using Stokes' theorem it can be written as,
\begin{equation}\label{SympPotentialWhitoutBoundary}
\Theta (\delta \phi^{A}) := \int_{\mathcal{M}} \d \theta ( \delta \phi^{A}) = \int_{\partial \mathcal{M}} \theta ( \delta \phi^{A})\, .
\end{equation}
This term is a $1-$form in the covariant phase space, namely, it acts on vectors $\delta \phi^{A}$ and returns a real number. Also it 
can be seen as a potential for the symplectic structure, as we shall see below. 
For such a reason, the term, $\Theta(\delta \phi^{A})$  is known as the \emph{symplectic potential} associated to a boundary $\partial\mathcal{M}$,  
and the integrand, $\theta ( \delta \phi^{A})$, is the \emph{symplectic potential current}\footnote{In the early literature, a symplectic potential is defined 
as an integral of $\theta$ over a spatial slice $M$, see, for example, \cite{abr}. Here, we are using the extended  definition of \cite{crv1} 
where it is important to consider the integral over the whole boundary $\partial\mathcal{M}$ in order to construct a symplectic structure.}.

Note that from Eqs. (\ref{VarActFormsWithoutBoundary}) and (\ref{SympPotentialWhitoutBoundary}), on the space of solutions defined by $E_{A} = 0$, 
the variation of the action becomes $\ed S(\delta) = \Theta (\delta \phi^{A})$.

As we pointed out in the previous section, the action (\ref{original_action}) may not be well defined, and one may need to introduce  a boundary term. 
In that case the well defined action becomes,
\begin{equation}\label{ActionFormsWithBoundary}
{\tilde S}[\phi^{A}] = \int_{\mathcal{M}}  (\mathbf{L} +  \d \varphi )\, ,
\end{equation}
where the boundary term in general depends on the fields, as well as their derivatives. 
Now, the variation has the form
\begin{equation}\label{GeneralVariatonForms}
\delta {\tilde S} = \int_{\mathcal{M}} E_{A} \wedge \delta \phi^{A} + \int_{\mathcal{M}} \d \left[\theta ( \delta \phi^{A}) +  \delta  \varphi \right].
\end{equation}
Note that we  can always add a term $\d Y$  to the symplectic potential current, $\theta \rightarrow \theta + \d Y$,
that will not change the corresponding symplectic potential. This object $Y$ can be seen as an intrinsic ambiguity of the formalism.
Thus, the most general symplectic potential can be written as,
\begin{equation}\label{GeneralTheta}
\tilde{\Theta} ( \delta) = 
\int_{\partial \mathcal{M}}  \left[\theta (\delta ) +  \delta  \varphi  + \d Y( \delta ) \right] =: \int_{\partial \mathcal{M}}  
\tilde{\theta} ( \delta ) ,
\end{equation}
where we have defined the extended symplectic potential current $\tilde\theta$.

Let us now take the exterior derivative of the symplectic potential, $\tilde{\Theta}(\delta \phi)$, acting on tangent vectors $\delta_{1}$ and 
$\delta_{2}$ at a point $\gamma$ of the CPS,
\begin{equation}\label{edTheta}
\ed \tilde{\Theta} (\delta_{1}, \delta_{2}) = \delta_{1} \tilde{\Theta} (\delta_{2}) -\delta_{2} \tilde{\Theta} (\delta_{1}) = 
2 \int_{\partial \mathcal{M}} \delta_{[1} \tilde{\theta} (\delta_{2]})\, .
\end{equation}
We can now define a space-time $3-$form, \emph{the symplectic current} $\tilde J(\delta_{1}, \delta_{2})$, to be the 
integrand of the RHS of (\ref{edTheta}),
\begin{equation}\label{DefJ}
\tilde J(\delta_{1}, \delta_{2}) := \delta_{1} \tilde{\theta}  (\delta_{2}) -\delta_{2} \tilde{\theta}  (\delta_{1})\, .
\end{equation}

The explicit form of the symplectic current is,
\begin{equation}\label{defJ}
\tilde{J}(\delta_{1}, \delta_{2}) = J(\delta_{1}, \delta_{2}) + 2\bigl(\delta_{[1} \delta_{2]} \varphi +  
 \delta_{[1}\d Y (\delta_{2]})\bigr)\, .
\end{equation}
where
\begin{equation}\label{defJ2}
J(\delta_{1}, \delta_{2}) := \delta_{1} \theta  (\delta_{2}) -\delta_{2} \theta  (\delta_{1})\, ,
\end{equation}
is the symplectic current associated to the action (\ref{original_action}).

Let us analyze  the terms of (\ref{defJ}). The term $ \delta_{[1} \delta_{2]} \varphi $ vanishes by antisymmetry, 
because $\delta_{1}$ and $\delta_{2}$ commute when acting on functions on CPS. 
Now, the last term of the RHS of (\ref{defJ}) can be written as $2\delta_{[1}\d Y (\delta_{2]}) = \d \chi (\delta_{1}, \delta_{2})$, where we have defined
$\chi (\delta_1,\delta_2) := 2\delta_{[1} Y (\delta_{2]} )$. We can do so given that 
$\d$ and $\delta_{i}$ commute. Since $\d$ and $\ed$ act on different spaces, the spacetime and the space of fields, respectively, they are independent. 
Thus, $\tilde{J}(\delta_{1}, \delta_{2})$ is given by
\begin{equation}\label{defJ1}
 \tilde{J}(\delta_{1}, \delta_{2}) = J(\delta_{1}, \delta_{2}) +  \d \chi (\delta_{1}, \delta_{2})\, .
\end{equation}
As we shall see later, the ambiguity in $\chi$ will be relevant in the examples that we consider below.

Therefore, one can see that, \emph{when one adds a boundary term to the original action it will not change the symplectic current}, and this result 
holds independently of the specific boundary conditions \cite{crv1}. 

Recall that, in the space of solutions, $\ed S (\delta) = \tilde{\Theta} (\delta)$, therefore from eqs. (\ref{edTheta}) and (\ref{DefJ}),
\begin{equation}
0 = \ed ^{2} S (\delta_{1}, \delta_{2}) =  \ed \tilde{\Theta} (\delta_{1}, \delta_{2})   =  
\int_{\mathcal{M}} \d \tilde J(\delta_{1}, \delta_{2}).
\end{equation}
Since we are integrating over \emph{any} region $\mathcal{M}$, it follows that $\tilde J$ is closed, i.e. $\d \tilde J = 0$. 
Note that $ \d\tilde J = \d (J+  \d \chi ) = \d J$ depends only on $\theta$, as can be seen from Eq. (\ref{defJ2}). If we now use Stokes' theorem, and 
select the orientation of $\partial\mathcal{M}$ as in Fig. \ref{regionM}, we have
\begin{equation}\label{intJzero}
0 = \int_{\mathcal{M}}  \d \tilde J(\delta_{1}, \delta_{2}) = 
\int_{\partial \mathcal{M}}   J(\delta_{1}, \delta_{2})  = \left(-  \int_{M_{1}} + \int_{M_{2}}  - \int_{\Delta}  + 
\int_{\mathcal{I}} \right)  J(\delta_{1}, \delta_{2}),
\end{equation}
where $\mathcal{M}$ is bounded by $\partial \mathcal{M} = M_{1} \cup M_{2} \cup  \Delta \cup \mathcal{I}$, $M_{1}$ and $M_{2}$ are 
space-like slices, 
$\Delta$ is an inner boundary and $\mathcal{I}=\lim_{\rho\to\infty}{\mathcal H}_\rho$ is an outer boundary at infinity.

Let us now consider the following three possible scenarios: 
First, consider the case when 
the asymptotic conditions ensures that the integral $\int_{\mathcal{I}}  J$ vanishes and  the boundary conditions (that might include no internal boundary) are such that $\int_{\Delta}  J$ also vanishes.
In this case, from (\ref{intJzero}) it follows
\begin{equation}
\int_{\partial \mathcal{M}}   J(\delta_{1}, \delta_{2})  =\left( - \int_{M_{1}} + \int_{M_{2}}   \right)  J(\delta_{1}, \delta_{2}) =0\, ,
\end{equation}
which implies that $\int_{M}  J$ is independent of the Cauchy surface. This allows us to define a \emph{conserved} pre-symplectic form over an arbitrary space like surface $M$,
\begin{equation}\label{symplectic_structure}
\bar{\Omega} (\delta_{1}, \delta_{2}) = \int_{M}  J  (\delta_{1}, \delta_{2})\, .
\end{equation}
By construction, the two form $\bar{\Omega}$ is closed, so it is justified to call it a (pre-)symplectic structure.
Note that in (\ref{intJzero}) there is only contribution from the symplectic current $J$, and not from the extended $\tilde{J}$ and, 
for that reason, the pre-symplectic
form does not depend on $\varphi$ (the contribution of the topological, total derivative, terms in the action) nor  on $\chi$ (the contribution
of total derivative terms in $\tilde{J}$). One should remark that the ambiguity in the definition of $\tilde{\theta}$ that we had pointed out, 
does not contribute to the pre-symplectic form. 

Next consider the case when  $\int_{\mathcal{I}}  J \neq 0$ (and the corresponding integral over $\Delta$ vanishes). 
Can we still define a conserved pre-symplectic form?  The answer is in the affirmative 
only if one can write 
\[
\int_{\mathcal{I}} J = \int_{\mathcal{I}} \d \beta = \int_{S_{\infty}^1} \beta - \int_{S_{\infty}^2} \beta\, ,
\]
 where 
$S_{\infty}^i = M_i \cap  \mathcal{I}$.
In this case we have,
\begin{equation}
0=\left(-  \int_{M_{1}} + \int_{M_{2}}  + \int_{\mathcal{I}} \right)  J = \left(-  \int_{M_{1}} + \int_{M_{2}}  \right)  J  +
\left(  \int_{S_{\infty}^{1}} - \int_{S_{\infty}^{2}} \right) \beta ,
\end{equation}
From which the corresponding conserved form is given by,
\be 
\bar{\Omega} (\delta_{1}, \delta_{2}) = \int_{M}  J  (\delta_{1}, \delta_{2}) - \int_{S_{\infty}} \beta (\delta_{1}, \delta_{2})\, .
\ee

Let us now consider the case when $\int_{\mathcal{I}}  J=0$, but we have a contribution from
an internal boundary. Then, let us consider the case when the integral $ \int_{\Delta} J$ may not vanish under the boundary conditions, 
as is the case with the  isolated horizon boundary conditions (more about this below). 
If, after imposing boundary conditions, we obtain that the pull back of the symplectic current on $\Delta$ is an exact form, $J|_{\Delta}=\d j$, then 
\begin{equation}\label{Jhorizon}
\int_{\Delta} J =  \int_{\Delta} \d j = \int_{\partial \Delta} j\, .
\end{equation}
Therefore we can define the \emph{conserved} pre-symplectic structure as,
\begin{equation}\label{sympl_struct}
\bar{\Omega}(\delta_{1}, \delta_{2})  = \int_{M}  J(\delta_{1}, \delta_{2})   +  \int_{S_\Delta} j (\delta_{1}, \delta_{2})\, ,
\end{equation}
where $S_\Delta = M\cap\Delta$.

Let us end this section by further commenting on the case when the symplectic current contains a total derivative \cite{crv1}, i.e. can be written  as
$\tilde{J}=J_0+\d\alpha$. Recall that, by our previous arguments, see (\ref{defJ1}) and (\ref{intJzero}), 
the $\d\alpha$ term does not appear in the symplectic structure. Therefore
it follows that, in the special case when $J_0=0$, the pre-symplectic structure is trivial $\bar\Omega =0$.
Nevertheless, in the literature, the symplectic structure is sometimes defined, from the beginning, as an integral of $\tilde{J}$
over a spatial hypersurface $M$.
Let us now describe the argument that one sometimes encounters in this context, in the simple case where $\tilde{J}=\d\alpha$. In this case one could 
{\it postulate} the existence of a pre-symplectic structure $\tilde\Omega_M$ as follows. Define
\begin{equation}\label{symplectic_structure_spatial}
 \tilde\Omega_M (\delta_1,\delta_2) := \int_M \d\alpha (\delta_1,\delta_2)\, ,
\end{equation}
therefore, from (\ref{intJzero}) the quantity $\tilde\Omega_M$ is independent on $M$ only if $\int_{\mathcal{I}} \d\alpha$ 
{\it and} $\int_{\Delta}\d\alpha$ vanish. In that case the object $\tilde\Omega_M$ is a conserved two-form that satisfies the 
definition of a pre-symplectic structure. It should be stressed though that such an object does {\emph{not}} follow from the systematic derivation we have introduced, starting from an action principle.

To summarize, in this part we have developed in detail the covariant Hamiltonian formalism in the presence of boundaries. As we have seen, 
there might be a contribution to the (pre-)symplectic structure coming from the boundaries. 
We have seen that the addition of boundary terms to the action does not modify the
conserved (pre-)symplectic structure of the theory, independently of the boundary conditions imposed.

\subsection{Symmetries}
\label{sec:3.2}

In this section we review how the covariant Hamiltonian formulation  addresses the existence of symmetries, and their associated conserved quantities. 
As a first step, let us recall the standard notion of a Hamiltonian vector field (HVF) in Hamiltonian dynamics.
A Hamiltonian vector field $Z$ is defined as a symmetry of the symplectic structure, namely 
\begin{equation}
\La_{Z} \Omega = 0.
\end{equation}
From this condition and the fact that $\ed \Omega = 0$ we have,
\begin{equation}\label{LieOmegaZero}
\pounds_{Z} \Omega = Z\cdot\ed \Omega + \ed (Z\cdot \Omega) = \ed (Z\cdot \Omega) = 0.
\end{equation}
where $Z\cdot\Omega\equiv i_{Z} \Omega$ is the \emph{contraction} of the 2-form $\Omega$ with the vector field $Z$. 
One can define the one-form $X_Z$ on $\Gamma$ as $X_Z (\delta): = (Z\cdot \Omega)(\delta)=\Omega (Z, \delta)$.
From the previous equation we  see that $X_Z$ is closed, that is, $\ed X_Z = 0 $. 
It follows from (\ref{LieOmegaZero}) and from the Poincar\'e lemma that locally (on the CPS), there exists a function $H_Z$ such that $X_Z = \ed H_Z$.
We call  $H_Z$ the \emph{Hamiltonian}  associated to $Z$, and is the function that generates the infinitesimal canonical transformation defined by $Z$. Furthermore, 
and by its own definition, $H_Z$ is a \emph{conserved quantity} along the flow generated by $Z$.
In what follows, we shall use in-distinctively the following notation for the directional derivative of any Hamiltonian $H$, 
along an arbitrary vector $\delta$: $X (\delta) = \ed H(\delta) = \delta H$.

Up to now the vector field $Z$ has been an arbitrary Hamiltonian vector field on $\Gamma$. Of special interest is the case
when one can relate it to certain spacetime symmetries. For instance, for field theories 
that possess a symmetry group, such as the Poincar\'e group on Minkowski spacetime, there will be 
Hamiltonian vector fields associated to the 
generators of the symmetry group. In this manuscript we are interested in exploring gravity theories that are diffeomorphism invariant. That is,  such that the diffeomorphism group on the spacetime manifold
acts as  (kinematical) symmetries of the action. 
Thus, it is particularly important to understand the role that these symmetries have in the 
Hamiltonian formulation. To be precise, 
one expects that diffeomorphisms play the role of {\em gauge} symmetries of the theory. However, it turns out that not all
diffeomorphisms  can be regarded as gauge. To distinguish them depends 
on the details of the theory, and is dictated by the properties of the corresponding Hamiltonian vector fields. Another
 important issue is to identify truly physical canonical transformations that
change the system. Those true motions could then be associated to symmetries of the theory. For instance, in the case of 
asymptotically flat spacetimes, {\em some} diffeomorphisms 
are regarded as gauge, while others represent nontrivial transformations at infinity and can be associated to the generators of the Poincar\'e group. 
In the case when the vector field $Z$ generates time evolution,  one expects $H_Z$ to be related to the energy, that is, the ADM energy at infinity. Other conserved, 
Hamiltonian charges can thus be found, and correspond to the generators of the asymptotic symmetries of the theory \cite{abr}.

In what follows we shall explore the aspects of the theory that allow us to separate the notion of gauge from standard symmetries of the theory.

\subsubsection{Gauge and degeneracy of the symplectic structure}

In the standard treatment of constrained systems, one starts out with the kinematical phase space $\Gamma_{\rm kin}$, 
and there exists a constrained surface $\bar\Gamma$ 
consisting of points that satisfy the constraints present in the theory. One then notices that the pullback of $\Omega$, 
the symplectic structure to $\bar\Gamma$  is degenerate 
(for first class constraints). These degenerate directions represent the gauge directions where two points are 
physically indistinguishable. In the covariant Hamiltonian 
formulation we are considering here, the starting point is the space $\Gamma$ of solutions to {\it all} the equations 
of motion, where a (pre-)symplectic structure is naturally 
defined, as we saw before. We call this a pre-symplectic structure since it might be degenerate.  
We say that $\bar{\Omega}$ is degenerate if there exist vectors $Z_i$ such that ${\bar{\Omega}}(Z_i, X) = 0$ for all  
$X$. We call $Z_i$ a degenerate direction (or an element of the kernel of $\bar{\Omega}$). If $\bar{\Omega}$ is 
degenerate we have a gauge system, with a gauge submanifold generated by the degenerate directions $Z_i$ 
(it is immediate to see that they satisfy the local integrability conditions to generate a submanifold).

Note that since we are on the space of solutions to the field equations, tangent vectors $X$ to $\Gamma$ must be solutions to the {\it linearized} equations of motion.
Since the degenerate directions $Z_{i}$ generate infinitesimal gauge transformations, configurations $\phi'$ and $\phi$ on  ${\Gamma}$, related by such transformations, are 
physically indistinguishable. That is, $\phi'  \sim \phi$ and, therefore, the quotient $\hat{\Gamma} = {\Gamma} / \sim$ constitutes the physical phase space of the system. It is 
only in the reduced phase space  $\hat{\Gamma}$ that one can define a non-degenerate symplectic structure $\Omega$.

In the next subsection we explain how vector fields are the infinitesimal generators of transformations on the space-time in general. Then we will point out when these 
transformations are diffeomorphisms and moreover, when these are also gauge symmetries of the system.

\subsubsection{Diffeomorphisms and gauge}

Let us start by recalling the standard notion of a diffeomorphism on the manifold ${\mathcal{M}}$. Later on, we shall see how, for diffeomorphism invariant theories, the induced 
action on phase space of certain diffeomorphisms becomes gauge transformations.

There is a one-to-one relation between vector fields on a manifold and families of transformations of the manifold onto itself. 
Let $\varphi$ be a one-parameter group of 
transformations on $\mathcal{M}$, the map $\varphi_{\tau}: \mathcal{M} \rightarrow \mathcal{M}$, defined by $\varphi_{\tau}(x) = \varphi(x,\tau)$, is a differentiable mapping. If 
$\mathbf{\xi}$ is the infinitesimal generator of $\varphi$ and $f \in C^{\infty}(\mathcal{M})$, $\varphi_{\tau}^{ *} f = f \circ \varphi_{\tau}$ also belongs to 
$C^{\infty}(\mathcal{M})$; 
then the Lie derivative of $f$ along $\xi$, $\pounds_{\mathbf{\xi}} f = \mathbf{ \xi}(f)$, represents the rate of change of the function $f$ under the family of 
transformations $\varphi_{\tau}$.
That is, the vector field $\xi$ is the generator of infinitesimal diffeomorphisms. Now, given such a vector field, a natural question is whether 
there exists a vector field 
$Z_\xi$ on the CPS that represents the induced action of the infinitesimal diffeos? As one can easily see, the answer is in the affirmative. 
 
In order to see that, 
let us go back a bit to Section~\ref{sec:2}. The action is defined on the space of histories (the space of all possible configurations) and, 
after taking the variation, the 
vectors $\delta \phi^{\alpha}$ lie on the tangent space to the space of histories. It is only after we restrict ourselves to the space of 
solutions $\Gamma$, that $\ed S (\delta) 
= \delta S = \Theta (\delta \phi^{A})$. Now these $\delta \phi^{A}$ represent \emph{any} vector on $T_{\phi^{A}} \Gamma$ 
(tangent space to $\Gamma$ at the point $\phi^{A}$). As we already mentioned, these $\delta \phi^{A}$ can be seen as ``small changes'' in the fields.
 What happens if we want the infinitesimal change of fields to be generated by a particular group of transformations (e.g. spatial translations, 
 boosts, rotations, etc)?
There is indeed a preferred tangent vector for the kind of theories we are considering. Given $\xi$, consider
\be
\delta_\xi \phi^{A}:=\pounds_\xi \phi^A\, .
\ee
From the geometric perspective, this is the natural candidate vector field to represent the induced action of infinitesimal diffeomorphisms on $\Gamma$. 
The first question is whether such objects are indeed
tangent vectors to $\Gamma$. It is easy to see that, {\it for kinematical diffeomorphism invariant theories, Lie derivatives satisfy the linearized equations of motion}.\footnote{
See, for instance \cite{W}. When the theory is not diffeomorphism invariant, such Lie derivatives are admissible vectors only when the defining vector field $\xi$ is a symmetry of 
the background 
spacetime.} 
Of course, in the presence of boundaries such vectors
must preserve the boundary conditions of the theory in order to be admissible (more about this below).
For instance, in the case of asymptotically flat boundary conditions, the allowed vector fields should
preserve the asymptotic conditions. 

Let us suppose that we have prescribed the phase space and pre-symplectic structure $\bar{\Omega}$, and a vector
field $\delta_\xi:=\pounds_\xi \phi^A$. The question we would like to pose is: when is such vector a degenerate direction of $\bar{\Omega}$? The equation 
that such vector $\delta_\xi$ must satisfy is then:
\be
\bar{\Omega}(\delta_\xi,\delta) = 0\, , \qquad \forall\; \delta\, . \label{cond-chida}
\ee
This equation will, as we shall see in detail below once we consider specific boundary conditions, impose some conditions on the behaviour of $\xi$ on the boundaries. An important 
signature of diffeomorphism invariant theories is that Eq.(\ref{cond-chida}) {\it only has contributions from the boundaries}. Thus, the vanishing of such terms will depend on the 
behaviour of $\xi$ there. In particular, if $\xi = 0$ on the boundary, the corresponding vector field is guaranteed to be a degenerate direction and therefore
to generate {\it gauge} transformations. In some instances, non vanishing vectors at the boundary also
satisfy Eq.~(\ref{cond-chida}) and therefore define gauge directions.

Let us now consider the case when $\xi$ is non vanishing on $\partial{\mathcal{M}}$ and Eq.~(\ref{cond-chida}) is not zero. In that case, we should have 
\be
\bar{\Omega}(\delta ,\delta_\xi )= \ed H_\xi (\delta) = \delta H_\xi\, ,\label{ecuacion-chida}
\ee
for some function $H_\xi$. This function will be the generator of the symplectic transformation generated by $\delta_\xi$. In other words, $H_\xi$ is the 
{\it Hamiltonian conserved charge} associated to the symmetry generated by $\xi$.

\noindent
Remark: One should make sure that Eq.~(\ref{ecuacion-chida}) is indeed well defined, given the degeneracy of $\bar{\Omega}$. In order to see that, 
note that one can add to $\delta_\xi$ an arbitrary `gauge vector' $Z$
and the result in the same: $\bar{\Omega}(\delta_\xi + Z,\delta)=\bar{\Omega}(\delta_\xi,\delta)$. Therefore, if such 
function $H_\xi$ exists (and we know that, locally, it does), it is insensitive to the existence of the gauge 
directions so it must be constant along those directions and, therefore, projectable to $\hat{\Gamma}$.
Thus, one can conclude that even when $H_\xi$ is defined through a degenerate pre-symplectic structure, it
is indeed a physical observable defined on the reduced phase space.

This concludes our review of the covariant phase space methods and the definition of gauge and  Hamiltonian conserved charges for diffeomorphism 
invariant theories. In the next part we shall revisit another aspect of symmetries on covariant theories, namely the existence of Noether conserved 
quantities, which are also associated to symmetries of field theories.

\subsection{Diffeomorphism invariance: Noether charge}
\label{sec:3.3}

In this part, we shall briefly review some results about  Noether conserved quantities and their relation to the Hamiltonian charges.  
For that, we shall rely on \cite{iw}. We know that to any Lagrangian theory 
invariant under diffeomorphisms we can associate 
a corresponding Noether current 3-form $J_N$. Consider infinitesimal diffeomorphism generated by 
a vector field $\xi$ on space-time. These diffeomorphisms
induce an infinitesimal change of fields, given by $\de_\xi\phi^A:=\La_\xi\phi^A$. 
From (\ref{VarActFormsWithoutBoundary}) it follows that the corresponding change in the 
lagrangian four-form is given by
\be
\La_\xi\Lb ={\rm E}_A\w\La_\xi\phi^A +\d\theta (\La_\xi\phi^A )\, .\label{3.1}
\ee
On the other hand, using Cartan's formula, we obtain
\be
\La_\xi\Lb = \xi\cdot\d\Lb + \d (\xi\cdot\Lb )=\d (\xi\cdot\Lb )\, ,\label{3.2}
\ee
since $\d\Lb =0$. From the previous equations we see that
\be 
{\rm E}_A\w\delta_\xi\phi^A+\d(\theta(\delta_\xi)- \xi\cdot\Lb)=0\, ,\label{3.3}
\ee
where  $\theta (\de_\xi ):=\theta (\La_\xi\phi^A )$. Now, we can define the {\it Noether current 3-form} as
\be
J_N (\de_\xi )=\theta (\de_\xi )-\xi\cdot\Lb\, .\label{noether}
\ee
From Eq. (\ref{3.3})  it follows that,
on the space of solutions, $\d J_N (\de_\xi )=0$, so at least locally one can define a
corresponding Noether charge density 2-form $Q_\xi$ (associated to $\xi$) as 
\be
J_N (\de_\xi ) = \d Q_\xi\, .\label{nch}
\ee
Following \cite{iw}, the integral of $Q_\xi$ over some compact surface $S$ is the Noether
charge of $S$ associated to $\xi$.
As we saw in the previous section the symplectic potential current $\theta$ is sensitive to the addition of an exact form, 
and a boundary term in the action principle, as seen in
(\ref{GeneralTheta}). In turn, that freedom translates into  ambiguities in the definition of $Q_\xi$. 
As we saw in section \ref{sec:3.1}, 
$\theta$ is defined up to an exact form: $\theta\to\theta +\d Y(\de )$. Also, the change in
Lagrangian $\Lb\to \Lb +\d\varphi$ produces the change $\theta\to\theta +\de\varphi$. 
As we have shown earlier these transformations leave invariant the symplectic structure, but 
they induce the following changes on the Noether current 3-form
\be\label{noether_current_change}
\tilde{J}_N(\de_\xi )= J_N(\de_\xi )+\d Y(\de_\xi )+\de_\xi\varphi -\xi\cdot\d\varphi\, ,
\ee
and the corresponding Noether charge 2-form becomes
\be\label{noether_charge_change}
\tilde{Q}_\xi = Q_\xi +Y(\de_\xi )+\xi\cdot\varphi +\d Z\, .
\ee
The last term in the previous expression is due to the ambiguity present in (\ref{nch}).

Let us see how one can obtain conserved quantities out of the Noether charge 2-form.
Since $\d {\tilde J}_N (\de_\xi )=0$ it follows, as in (\ref{intJzero}), that 
\begin{equation}
 0 = \int_{\mathcal{M}}  \d {\tilde J}_N (\de_\xi )   = \int_{\partial \mathcal{M}} {\tilde J}_N (\de_\xi )  
 = \left(-  \int_{M_{1}} + \int_{M_{2}}  - \int_{\Delta}  + \int_{\mathcal{I}} \right){\tilde J}_N (\de_\xi ),
\end{equation}
and we see that if $\int_{\Delta}{\tilde J}_N (\de_\xi )=\int_{\mathcal{I}}{\tilde J}_N (\de_\xi )=0$ then the
previous expression implies the existence of the conserved quantity (independent on the choice
of $M$),
\begin{equation}
 \int_{M}{\tilde J}_N (\de_\xi )=\int_{\partial M}{\tilde Q}_\xi \, .
\end{equation}
Note that the above results are valid only on shell. If the corresponding integrals of $\tilde{J}_N(\de_\xi)$ do not vanish 
on the boundaries, one has to proceed with care.

In the covariant phase space, and for $\xi$ arbitrary and fixed, we have \cite{iw}
\be
\de J_N (\de_\xi )=\de \theta (\de_\xi )-\xi\cdot \de\Lb =\de \theta (\de_\xi )-\xi\cdot\d\theta (\de )\, .
\ee
Since, $\xi\cdot\d\theta =\La_\xi \theta -\d (\xi\cdot\theta )$ and
$\de \theta (\de_\xi )-\La_\xi \theta (\de )=J(\de ,\de_\xi )$ by the definition of the 
symplectic current $J$ (\ref{DefJ}), it follows that the relation between the symplectic current $J$
and the Noether current 3-form $J_N$ is given by
\be
J(\de ,\de_\xi )=\de J_N (\de_\xi )-\d (\xi\cdot\theta (\de ))\, .\label{nc}
\ee
We shall use this relation in the following sections, for the various actions that describe
first order general relativity, to clarify the relation between the Hamiltonian and Noether charges.
As was shown explicitly in \cite{crv1}, in general, a Noether charge does not 
correspond to a Hamiltonian charge generating symmetries of the phase space. 

\section{The action for gravity in the first order formalism}
\label{sec:4}

In this manuscript, we are interested in the most general action for four-dimensional general relativity in the first order formalism. 
In this section we shall analyze the variational principle, and we shall focus on the contribution coming from
each of the allowed terms. In first order gravity, the choice of basic variables is the following:
A pair of co-tetrads $e_a^I$ and a Lorentz $SO(3,1)$ connection $\o_{aIJ}$ on the spacetime $\mathcal{M}$, possibly with 
boundary. In order 
for the action to be physically relevant, it should reproduce the equations of motion for general relativity and
be: 1) differentiable, 2) finite on the configurations with a given asymptotic behaviour if the spacetime is unbounded, and
3) invariant under diffeomorphisms and local internal Lorentz transformations. 

The most general action that gives the desired
equations of motion and is compatible with the symmetries of the theory is given by a linear combination of the
Palatini action, $S_{\mathrm{P}}$, Holst term, $S_{\mathrm{H}}$, and three topological terms, Pontryagin, 
$S_\Po$, Euler, $S_\Eu$, and Nieh-Yan, $S_\ny$, invariants \cite{perez}. 
Therefore the complete action can be written as,
\begin{equation}\label{complete-action}
S[e, \omega] = S_{\rm P} + \alpha_{1} S_{\rm Po} + 
\alpha_{2}S_{\rm E} + S_{\rm H} + \alpha_{3} S_{\rm NY} + S_{\rm BT}\, .
\end{equation}
Here $\alpha_{i}$, $i=1,2,3$, are arbitrary coupling constants, and $S_{\rm BT}$ represents all boundary terms that need to be added.
As we shall see, the Palatini term contains
the information of the Einstein-Hilbert (2nd order) action, in the sense that, for spacetimes without boundaries, both actions 
are well defined and yield the same equations of motion. Thus, the Palatini term represents the backbone of
the formalism. One of the question that we want to address is that of the contribution to the formalism coming from 
the various additional terms in the action. 
Since we are considering a spacetime region $\mathcal{M}$ with boundaries, one should pay special attention 
to  the boundary conditions. For instance, it turns out that the Palatini action, as well as
Holst and Nieh-Yan terms are not differentiable for asymptotically flat spacetimes, and 
appropriate boundary terms should be provided (see more in the next section). 

This section has four parts, where we are going to analyze, one by one, 
all of the terms of the most general action (\ref{complete-action}). We shall take the corresponding variation
of the terms and identify both their contributions to the equations of motion and to the symplectic current. 
Since we are not considering yet any particular boundary conditions, the results of this section are universal.

\subsection{Palatini action}
\label{sec:4.1}

Let us start by considering the Palatini action without a boundary term, 
\begin{equation}
  S_{\mathrm{P}} = - \frac{1}{2 \kappa} \int_{\mathcal{M}} \Sigma^{IJ} \wedge  F_{IJ}  \, , 
  \label{PalatininoBoundary}
\end{equation}
where $\k =8\pi G$, 
$\Sigma^{IJ} = \star (e^I\w e^J):=\frac{1}{2} \epsilon^{IJ}\,_{JK}e^{J}\wedge e^{K}$,
$F_{IJ}=\d\o_{IJ}+\o_{IK}\w{\o^K}_J$ is a curvature two-form of the connection $\o$
and, as before, $\partial \mathcal{M} = M_{1} \cup M_{2} \cup  \Delta \cup \mathcal{I}$.
If one varies this action, the boundary term that one gets is proportional to $\int_{\partial\mathcal{M}} \Sigma^{IJ}\wedge
\delta \o_{IJ}$. Of course, the differentiability of the action depends on the details of the boundary conditions. 
If these were such that the previous term vanishes, then one would not need to introduce any further term to make the 
action differentiable. Unfortunately, in most situations of interest, this is not the case. In many instances, one would like to
fix some variations of the boundary metric, which implies fixing certain components of the tetrad $e_a^J$. It is then 
costumary to add boundary terms to the original action that modify the resulting boundary term after the variation of the action. 
Let us now review some of these choices. 

The simplest choice is to take the boundary term 
\be
S_{\textrm{B}} = \frac{1}{2 \kappa} \int _{\partial \mathcal{M}} \Sigma^{IJ} \wedge \omega_{IJ}\, .
\label{boundary-1}
\ee
If we now vary the Palatini action (\ref{PalatininoBoundary}) together with the boundary term (\ref{boundary-1}), the resulting 
contribution from the boundary is now of the form $\int_{\partial\mathcal{M}} \delta\Sigma^{IJ}\wedge
 \o_{IJ}$, which is what we wanted. Under appropriate boundary conditions imposed now on $\delta e_a^J$, the complete action 
 becomes differentiable if, for instance, one fixes $e_a^J$ on the boundary, or the falloff conditions for an unbounded
 region $\mathcal{M}$ are strong enough to cancel the term. As we have remarked before, one should also ensure that the action 
 together with the boundary term is finite.

Note that the boundary term (\ref{boundary-1}) is not manifestly gauge invariant, but, for the appropriate boundary  conditions, 
this might not be a problem. For instance, for asymptotically flat and AdS boundary conditions,
as pointed out in \cite{aes}, it is effectively
gauge invariant on the spacelike surfaces $M_1$ and $M_2$ and also in the asymptotic region $\mathcal{I}$.
This is due to the fact that the only allowed gauge transformations that preserve the asymptotic conditions 
are such that the boundary terms remain invariant. To see that,
let us first consider  the behaviour of this boundary term on $M_1$ (or $M_2$). First we ask that the compatibility
condition between the co-tetrad and connection should be satisfied on the boundary. Then, we partially
fix the gauge on $M$, by fixing the internal time-like tetrad $n^I$, such that $\partial_a n^I=0$ and
we restrict field configurations such that $n^a=e^a_In^I$ is the unit normal to $M_1$ and $M_2$. Under
these conditions it has been shown in \cite{aes} that on $M$, $\Sigma^{IJ} \wedge  \omega_{IJ}=2K\d^3V$,
where $K$ is the trace of the extrinsic curvature of $M$. Note that this is 
the Gibbons-Hawking surface term that is needed in the Einstein-Hilbert action, with the constant 
boundary term equal to zero. On the other hand, at spatial infinity, $\mathcal{I}$, we fix the
co-tetrads and only permit gauge transformations that reduce to  the identity at infinity. Under these conditions the
boundary term is gauge invariant at $M_1$, $M_2$ and $\mathcal{I}$. As we shall show later the term  is also
invariant under the residual local Lorentz transformations at a weakly isolated horizon, when such a
boundary exists. 

It is important to note that
there are other proposals for manifestly gauge invariant boundary terms for the  Palatini action, as for example those 
introduced in  \cite{bn}, \cite{obukhov} and \cite{aros}.

Let us first recall the boundary term put forward in \cite{aros}. The idea is to substract a second boundary term with a 
fixed connection $\o_0$ such that it has the form,
\be
S_{\textrm{BA}} = \frac{1}{2 \kappa} \int _{\partial \mathcal{M}} \Sigma^{IJ} \wedge (\omega_{IJ} - \o_{0IJ})\, .
\label{boundary-2}
\ee
Note that this is manifestly gauge invariant since the difference of two connections is a tensorial object.
This term was introduced to make the action finite, in analogy with the Gibbons-Hawking-York term in second order gravity.

The next proposal that we want to consider was constructed with the purpose of having a well defined first order action 
when there is a boundary, in the context of  Einstein-Cartan theory \cite{obukhov}. Here the idea is to add a term that contains the
covariant derivative of the normal to the boundary. This proposal was extended in \cite{bn}, where
the following boundary term was introduced,
\begin{equation}
S_{\textrm{BN}}:=-\frac{1}{\kappa}\int _{\partial \mathcal{M}}\, \frac{1}{\tilde{n}\cdot \tilde{n}}\, \Sigma^{IJ} 
\wedge \tilde{n}_I D\tilde{n}_J\, , \label{Palatini_BN_boundary}
\end{equation}
where $\tilde{n}_I$ is a non-unit co-normal, defined as $\frac{\tilde{n}_I}{\sqrt{\tilde{n}\cdot \tilde{n}}}:=
sr^a e_{aI} = sr_I$, where $r^a$ is a unit normal to $\partial \mathcal{M}$, $s=r^ar_a$ and 
$D\tilde{n}_J = \d \tilde{n}_J + {\omega_J}^K \tilde{n}_K$. (The choice of a non-unit normal was made since the authors wanted 
to study the signature change along the boundary.)
This term is obtained without imposing the time gauge condition and is equivalent to Gibbons-Hawking term
(under the half on-shell condition $De^I=0$).
It is manifestly gauge invariant and well defined for finite boundaries, but for example, it is not well defined
for asymptotically flat spacetimes. In time gauge it reduces to (\ref{boundary-1}), since
\begin{equation}
S_{{\textrm{BN}}}=\frac{1}{2\kappa} \int _{\partial \mathcal{M}} \Sigma^{IJ} \wedge \bigl( \omega_{IJ}-
\frac{2\tilde{n}_I\d \tilde{n}_J}{\tilde{n}\cdot \tilde{n}}\bigr) \, .
\end{equation}

From all these possibilities, we shall restrict ourselves in what follows, to the simplest case considered above, namely, the action 
\begin{equation}
  S_{\mathrm{PB}} = - \frac{1}{2 \kappa} \int_{\mathcal{M}} \Sigma^{IJ} \wedge  F_{IJ}  +   
  \frac{1}{2 \kappa} \int _{\partial \mathcal{M}} \Sigma^{IJ} \wedge \omega_{IJ}\, .
  \label{PalatiniplusBoundary}
\end{equation}
We are making this choice because, for asymptotically flat falloff conditions, the boundary term is gauge invariant, 
as discussed above, and the total action is finite and differentiable.

The variation of (\ref{PalatiniplusBoundary}) is,
\begin{eqnarray}
\delta S_{\mathrm{PB}}
= - \frac{1}{2 \kappa} \int_{\mathcal{M}} \left[ \varepsilon^{IJ}\, _{KL} 
\delta e^{K} \wedge e^{L} \wedge F_{IJ} - D\Sigma_{IJ} \wedge \delta \omega^{IJ} - 
\d (\delta \Sigma^{IJ} \wedge \omega_{IJ}  ) \right]\label{variationPalatini}\, ,
\end{eqnarray}
where 
\begin{equation}\label{InternalCovDev}
 D \Sigma_{IJ} = \d \Sigma_{IJ} - \omega_{I}\, ^{K} \wedge \Sigma_{KJ} + \omega_{J}\, ^{K} \wedge \Sigma_{KI}\, . 
 \end{equation}
We shall show later that the contribution of the boundary term $\delta \Sigma^{IJ} \wedge \omega_{IJ}$ vanishes at $\mathcal{I}$ and $\D$, 
so that from (\ref{variationPalatini}) we obtain the following equations of motion
\begin{eqnarray}
\label{EOMPalatiniFe} 
\varepsilon_{IJKL} e^{J} \wedge F^{KL} &=& 0\, , \\
\label{EOMPalatiniDe} \varepsilon_{IJKL} e^{K} \wedge D e^{L} &=& 0\, ,
\end{eqnarray}
where $T^L:=De^L=\d e^L +{\o^L}_K \w e^K$ is the torsion two-form. 
From (\ref{EOMPalatiniDe}) it follows that $T^L = 0$, and this is the 
condition for the compatibility of $\omega_{IJ}$ and $e^{I}$, that implies
\be
\o_{aIJ}=e^b_{[I}\partial_a e_{bJ]}+\Gamma^c_{ab}e_{c[I}e^b_{J]}\, ,
\ee
where $\Gamma^c_{ab}$ are the Christoffel symbols of the metric $g_{ab}=\eta_{IJ}e_a^Ie_b^J$.
Now, the equations (\ref{EOMPalatiniFe}) are equivalent to Einstein's equations $G_{ab}=0$ \cite{romano}.

From the equation (\ref{variationPalatini}), the symplectic potential for $S_{\mathrm{PB}}$ 
is given by
\begin{equation}\label{SympPotentialPalatini}
\Theta_{\mathrm{PB}} (\delta ) = 
 \frac{1}{2 \kappa} \int_{\partial \mathcal{M}} \delta \Sigma^{IJ} \wedge \omega_{IJ}\, .
\end{equation}
Therefore from (\ref{defJ}) and (\ref{SympPotentialPalatini}) the corresponding symplectic current is,
\begin{equation}\label{JPalatini}
J_{\mathrm{P}}(\delta_{1}, \delta_{2})  = -\frac{1}{2 \kappa} \left( \delta_{1} \Sigma^{IJ} \wedge \delta_{2} \omega_{IJ} 
- \delta_{2} \Sigma^{IJ} \wedge \delta_{1} \omega_{IJ} \right)\, .
\end{equation}
As we discussed in Sec.~\ref{sec:3},  the symplectic current  
is insensitive to the boundary term in the action.

As we shall discuss in the following sections, the Palatini action, in the asymptotically flat case, is not well  defined, 
but it can be made differentiable and finite after the addition of the corresponding 
boundary term already discussed \cite{aes}. 
Furthermore, we shall also show that in the case when the spacetime has as internal 
boundary an isolated horizon, the contribution at the horizon to the variation of the Palatini action, either with a boundary 
term \cite{cg} or without it \cite{afk}, vanishes. 

\subsection{Holst and Nieh-Yan terms}
\label{sec:4.2}

The first additional term to the gravitational action that we shall consider is the so called {\it Holst} term \cite{holst}, 
first introduced with the aim of having a variational 
principle whose $3+1$ decomposition yielded general relativity in the  Ashtekar-Barbero (real) variables \cite{barbero}.
It turns out that the Holst term, when added to the Palatini action, does not change the equations of motion
(although it is not a topological term), so that in the Hamiltonian formalism its addition corresponds to
a canonical transformation. This transformation leads to the Ashtekar-Barbero variables that are the basic
ingredients in the loop quantum gravity approach. The Holst term is of the form
\begin{equation}\label{Holstterm}
S _{\mathrm{H}}= - \frac{1}{2 \kappa\g} \int_{\mathcal{M}} \Sigma^{IJ}\w \star F_{IJ} \, ,
\end{equation}
where $\g$ is the Barbero-Immirzi parameter. As we shall show in the next section, the Holst term is finite but 
not differentiable for asymptotically flat spacetimes, so  an appropriate boundary term should be added in order 
to make it well defined. 

A boundary term that makes the Holst term differentiable  was proposed in \cite{cwe}, and it is of the form
\begin{equation}\label{Holst-boundary-1}
 S _{\mathrm{BH}}=\frac{1}{2 \kappa\g} \int _{\partial \mathcal{M}}  \Sigma^{IJ} \wedge \star\o_{IJ} \, ,
\end{equation}
as an analogue of the boundary term (\ref{boundary-1}) for the Palatini action.  Then, we define 
$S_{\mathrm{HB}}=S_{\mathrm{H}}+S_{\mathrm{BH}}$, such that
\begin{equation}\label{HolsttermBoundary}
S_{\mathrm{HB}}=- \frac{1}{2 \kappa\g} \int_{\mathcal{M}} \Sigma^{IJ}\w \star F_{IJ}
+\frac{1}{2 \kappa\g} \int _{\partial \mathcal{M}}  \Sigma^{IJ} \wedge \star\o_{IJ}\, .
\end{equation}
The variation of $S_{\mathrm{HB}}$ is given by
\begin{equation}\label{variationHolst}
\de  S _{\mathrm{HB}}= - \frac{1}{2 \kappa\g} \int_{\mathcal{M}} 2F_{IJ}\w e^I\w \de e^J +
D\S^{IJ}\w\star (\de\o_{IJ}) -\d (\de\S^{IJ}\w\star\o_{IJ})\, ,
\end{equation}
and it leads to the following equations of motion in the bulk: $D\S^{IJ}=0$ and $e^I\w F_{IJ}=0$. 
The second one is just the Bianchi identity, and we see that the Holst term does not modify the equations of motion of
the Palatini action. The contribution of the boundary term (that appears in the variation) should vanish
at $\mathcal{I}$ and $\D$, in order to have a well posed variational principle.
In the following section we shall see that this is indeed the case for the boundary conditions considered there.

On the other hand we should also examine the gauge invariance of the boundary term  (\ref{Holst-boundary-1}).
Using the equation of motion $De^I=0$, on the Cauchy surface $M$, we obtain
\be
\int_{M}  \Sigma^{IJ} \wedge \star\o_{IJ}=
\int_{M} e^I\w\d e_I = 4\int_{M} \epsilon^{abcd}r_a e^I_c\partial_d e_{bI}\sqrt{h}\d^3x\, .
\ee
where $h$ is the determinant of the induced metric on $M$, $r^a$ is the unit normal to $M$ and
$\epsilon^{abcd}$ is the Levi Civita tensor.
It follows that this term is not gauge invariant at $M$.
As we shall see in the following section, at the asymptotic region 
it is gauge invariant, and also  at $\D$. In the analysis of differentiability of the action and
the construction of the symplectic structure and conserved quantities there is no contribution from
the spacial surfaces $M_1$ and $M_2$, and we can argue that the non-invariance of the boundary term 
(\ref{Holst-boundary-1}) is not important, but it would be desirable to have a boundary term that is
compatible with all the symmetries of the theory. 

Let us now consider another choice for a boundary term for the Holst term, that has an advantage to be 
manifestly gauge invariant. This boundary term was proposed in \cite{bn}, and is proportional to
the Nieh-Yan topological invariant, $S_{\ny}$. This topological term 
is related to the torsion $T^I:=De^I$, and is of the form \cite{nieh-yan,nieh},
\begin{equation}\label{Nieh-Yan}
S_{\ny} = \int_{\mathcal{M}} \left( D e^{I} \wedge D e_{I}  - \Sigma^{IJ} \wedge \star F_{IJ} \right)= 
\int_{\partial \mathcal{M}} De^{I} \wedge e_{I}\, .
\end{equation}
Note that the Nieh-Yan term can be written as
\begin{equation}\label{relationNYHolst}
S_{\ny} = 2\k\g S_{\mathrm{H}}+\int_{\mathcal{M}}  D e^{I} \w D e_{I}\, .
\end{equation}
In the next section we shall show that the Nieh-Yan term is finite, but not differentiable, for 
asymptotically flat spacetimes, in such a way that
the surface term in the variation of Neih-Yan term cancels the surface term in the variation of the 
Holst term. As a result, we can add the Neih-Yan topological invariant as a boundary term to the Holst term and define 
\be  \label{hny}
S_{\mathrm{HNY}}:=S_{\mathrm{H}}-\frac{1}{2\k\g}S_{\ny}=
-\frac{1}{2 \kappa\g}\int_{\mathcal{M}}
De^I\w De_I\, ,
\ee
which turns out to be well defined, finite and manifestly gauge invariant for the boundary conditions
that we will consider in the next sections. 

In what follows we shall consider the properties of both terms, $S_{\mathrm{HB}}$ and $S_{\mathrm{HNY}}$.
As we mentioned earlier, the first choice, $S_{\mathrm{HB}}$, is convenient for the introduction of Ashtekar-Barbero
variables in the canonical Hamiltonian approach, while the second one, $S_{\mathrm{HNY}}$, is more appropiate
in the presence of fermions when one has spacetimes with torsion. In that case, as shown in \cite{mercuri}, one 
should consider the Neih-Yan topological term, instead of the Holst term.
Since, as we shall see,  the Neih-Yan term is not well defined for our boundary conditions, one should 
consider the term $S_{\mathrm{HNY}}$ instead. 

To end this section, let us calculate the symplectic potential for $S_{\mathrm{HB}}$ and $S_{\mathrm{HNY}}$.
It is easy to see that the symplectic potential for $S_{\mathrm{HB}}$ is given by \cite{cwe}
\be
\Theta_{\mathrm{HB}} (\de )=\f {1}{2\k\g}\int_{\partial \mathcal{M}} \de\S^{IJ}\w\star\o_{IJ}=
\f {1}{\k\g}\int_{\partial \mathcal{M}} \de e^I\w\d e_I\, ,
\ee
where in the second line we used the equation of motion $De^I=0$. The symplectic current is given by
\begin{equation}\label{symplectic_current_holst}
J_{\mathrm{HB}}(\delta_{1}, \delta_{2})  = \frac{1}{\k\g}\d\, (\de_1 e^I\w\de_2 e_I)\, .
\end{equation}
As we have seen in the subsection \ref{sec:3.1}, when the symplectic current is a total derivative, the covariant
Hamiltonian formalism indicates that the corresponding (pre)-symplectic structure vanishes.
As we also remarked, one could in principle postulate a conserved two form $\tilde\O$
if $\int_{\mathcal I}J_{\mathrm{H}}=0$ and $\int_{\D}J_{\mathrm{H}}=0$, in which case
this term defines a conserved symplectic structure. We shall, for completeness, consider this
possibility in Sec.~\ref{sec:6}, after the appropriate boundary conditions have been introduced. 

On the other hand, the symplectic potential for $S_{\mathrm{HNY}}$ is given by
\begin{equation}\label{HNY_symplectic_potential}
\Theta_{\mathrm{HNY}} (\de )=-\f {1}{\k\g}\int_{\partial \mathcal{M}} De^I\w\de e_I =0\, .
\end{equation}
We see that in this case the symplectic potential vanishes.


\subsection{Pontryagin and Euler terms}
\label{sec:4.3}

As we have seen before, in four spacetime dimensions there are three topological invariants constructed 
from $e^I$, $F_{IJ}$ and $D e^I$, consistent with diffeomorphism and local Lorentz invariance. They are all 
exact forms and therefore, do not contribute to the equations of motion. Nevertheless, they should be finite and 
their variation  on the boundary of the spacetime region ${\mathcal{M}}$ 
should vanish. Apart from the Neih-Yan term  that we have considered 
in the previous section, there are also the Pontryagin and Euler terms that are constructed solely from
the curvature $F_{IJ}$ and its dual (in the internal space) $\star F_{IJ}$.

These topological invariants can be thought of as 4-dimensional Lagrangian densities
 defined on a manifold $\mathcal{M}$, that additionally are exact forms,  but they can also 
be seen as terms living on $\partial \mathcal{M}$. In that case it is obvious that they do not contribute
to the equations of motion in the bulk. But a natural question may arise. If we take the Lagrangian 
density in the bulk and take the variation, what are the corresponding equations of motion
in the bulk? One can check that, 
for Pontryagin and Euler, the resulting equations of motion are trivial in the sense that one only gets 
the Bianchi identities, while for the Nieh-Yan term they vanish identically.
Let us now see how each of this terms contribute to the variation of the action.

The action corresponding to the Pontryagin term is given by,
\be
\label{Pontryagin} S_{\Po} =\int_{\mathcal{M}} F^{IJ} \wedge F_{IJ} = 
2\int_{\partial \mathcal{M}} \left( \omega_{IJ} \wedge d\omega^{IJ} + 
\frac{2}{3} \omega_{IJ}\wedge \omega^{IK} \wedge \omega_{K}\,^{J} \right)\, .
\ee
The boundary term  is the $SO(3,1)$ Chern-Simons Lagrangian density, $L_{\textrm{CS}}\,$.
We can either view the Pontryagin term as a bulk term or as a boundary term and the derivation of the 
symplectic structure in either case should render equivalent descriptions.
The variation of $S_{\Po}$, calculated from the LHS expression in (\ref{Pontryagin}), is
\be
\de S_{\Po}=-2 \int_{\mathcal{M}} DF^{IJ}\w\de\o_{IJ}+
2\int_{\partial \mathcal{M}}F^{IJ}\w\de\o_{IJ}\, .\label{var_Po1}
\ee
We can then see that it does not contribute to the equations of motion in the bulk, due to the Bianchi
identity $DF^{IJ}=0$. Additionally, the surface integral in
(\ref{var_Po1}) should vanish for the variational principle to be well defined. 
We will show in later sections that this is indeed the case for boundary conditions of interest to us, namely, asymptotically 
flat spacetimes possibly with an isolated horizon as inner boundary. In this case, the corresponding symplectic current is 
\begin{equation}
J_{\Po}^{\rm bulk}(\de_1,\de_2)=2(\de_1 F^{IJ}\w\de_2\o_{IJ} -  \de_2 F^{IJ}\w\de_1\o_{IJ})\, .
\end{equation}

Let us now consider  the variation of the Pontryagin term directly from the 
RHS of (\ref{Pontryagin}), where it is a boundary term. We obtain
\begin{equation}\label{var_Po2}
\de S_{\Po}=2 \int_{\partial \mathcal{M}}\de L_{\textrm{CS}}\, .
\end{equation}
One should expect the two expressions for $\de S_{\Po}$ to be identical. This is indeed the case since
$F^{IJ}\w\de\o_{IJ}=\de L_{CS}+\d (\o^{IJ}\w\de\o_{IJ})$. 
The first expression
(\ref{var_Po1}) is more suited for the analysis of the differentiability of the Pontryagin term, 
but from the second one (\ref{var_Po2}), the vanishing of the  symplectic current is more apparent, since
\begin{equation}
J_{\Po}^{\rm bound}(\de_1,\de_2)=4\,\de_{[2}\de_{1]} L_{\textrm{CS}}=0\, .
\end{equation}
Note that, at first sight it would seem that there is an ambiguity in the definition
of the symplectic current that could lead to different symplectic structures. 
Since the relation between them is given by
\begin{equation}
 J_{\Po}^{\rm bulk}(\de_1,\de_2)=J_{\Po}^{\rm bound}(\de_1,\de_2)+4\,\d (\de_2\o^{IJ}\w\de_1\o_{IJ})\, ,
\end{equation}
it follows that $J_{\Po}^{\rm bulk}(\de_1,\de_2)$ is a total derivative, that does not contribute in
(\ref{intJzero}), and from the systematic derivation of the symplectic structure described in \ref{sec:3.1}, 
we have to conclude that it does not contribute to the symplectic structure. 
This is consistent with the fact that $J_{\Po}^{\rm bound}$
and $J_{\Po}^{\rm bulk}$ correspond to the same action. As we have remarked in  Sec.~\ref{sec:3}, a total
derivative term in $J$, under some circumstances, could be seen as generating a non-trivial symplectic
structure $\tilde\O$ on the boundary of $M$. But the important thing to note here is that if one were to 
introduce such object $\tilde\O$, one would run into an inconsistency, given that one would arrive to
two distinct pre-symplectic structures for the same action.
 Thus, consistency of the formalism requires that $\tilde\O=0$.

Let us now consider the action for the Euler term, which is given by,
\be
\label{Euler} S_{\Eu} = \int_{\mathcal{M}} F^{IJ} \wedge \star F_{IJ}= 
2\int_{\partial \mathcal{M}} \left( \star \omega_{IJ} \wedge \d\omega^{IJ} + 
\frac{2}{3} \star{\omega}_{IJ}\wedge \omega^{IK} \wedge \omega_{K}\,^{J} \right)\, ,
\ee
whose variation, calculated from the expression in the bulk, given by
\be
\de S_{\Eu}= -2 \int_{\mathcal{M}} \star DF^{IJ}\w\de\o_{IJ}+
2 \int_{\partial \mathcal{M}}\star F^{IJ}\w\de\o_{IJ}\, .\label{var_Eu}
\ee
Again, the action will only be well defined if the boundary contribution to the
variation (\ref{var_Eu}) vanishes.
In the following section we shall see that it indeed vanishes for our boundary conditions. 
Let us denote by $L_{\textrm{CSE}}$ the boundary term on the RHS of (\ref{Euler}), then
we can calculate the variation of $S_{\Eu}$ from this term directly as
\be
\de S_{\Eu}=2 \int_{\partial \mathcal{M}}\de L_{\textrm{CSE}}\, .
\ee
Finally, as before, the corresponding contribution from the Euler term to the symplectic current vanishes.

\section{Boundary conditions}
\label{sec:5}

We have considered the most general action for general relativity in the first order formalism, including boundaries, 
in order to have a well defined action principle and 
covariant Hamiltonian formalism. We have left, until now, the boundary conditions unspecified, other that assuming that 
there is an outer and a possible inner boundary to the 
region ${\mathcal M}$ under consideration. In this section we shall consider boundary conditions that are physically 
motivated: 
asymptotically flat boundary conditions that capture the notion of isolated systems and, for the inner boundary, 
isolated horizons boundary conditions. In this way, we 
allow for the possibility of spacetimes that contain a black hole. This section has two parts. In the first one, we consider 
the outer boundary conditions and in the second part, 
the inner horizon boundary condition. In each case, we study the finiteness of the action, its variation and its differentiability. 
Since this manuscript is to be self-contained, we include a detailed discussion of the boundary conditions before analyzing the 
different contributions to the action.

\subsection{Asymptotically flat spacetimes}
\label{sec:5.1}

In this part, we are interested in spacetimes that at infinity resemble  flat spacetime.
That is,  the spacetime metric approaches a Minkowski metric at infinity 
(in some appropriately chosen coordinates). 
Here we shall review the standard definition of asymptotically flat spacetimes in the first order formalism 
(see e.g. \cite{aes}, \cite{cwe} and for a nice and pedagogical introduction in the metric formulation \cite{abr} and \cite{W}). 
Here we give a brief introduction into asymptotically flat spacetimes, following closely \cite{aes}.

In order to describe the behaviour of the metric at spatial infinity, we will focus on the region 
$\mathcal{R}$, that is the region outside the light cone of some point $p$.  We define a $4-$dimensional radial 
coordinate $\rho$ given by $\rho^{2} = \eta_{ab} x^{a} x^{b}$, where $x^{a}$ are the Cartesian coordinates 
of the Minkowski metric $\eta$ on $\mathbb{R}^{4}$ with origin at $p$. 
We will foliate the asymptotic region by timelike hyperboloids, $\mathcal{H}_\rho$, given by $\rho = \mathrm{const}$, 
that lie in $\mathcal{R}$. 
Spatial infinity $\mathcal{I}$ corresponds to a limiting hyperboloid when $\rho\to\infty$.
The standard angular coordinates on a hyperboloid are denoted by $\Phi^i =(\chi, \theta, \phi )$,
and the relation between Cartesian and hyperbolic coordinates is given by:
$x(\rho, \chi, \theta, \phi)= \rho \cosh \chi \sin \theta \cos \phi, \,\,\,\,\, y(\rho, \chi, \theta, \phi) 
= \rho \cosh \chi \sin \theta \sin \phi, \,\,\, z(\rho, \chi, \theta, \phi) = 
\rho \cosh \chi \cos \theta, \,\,\, t(\rho, \chi, \theta, \phi) = \rho \sinh \chi$.

We shall consider functions $f$ that \emph{admit an asymptotic expansion to order $m$ } of the form,
\begin{equation}
f(\rho, \Phi) = \sum_{n=0} ^{m} \frac{\,^{n}f(\Phi)}{\rho^{n}} + o(\rho^{-m}),
\end{equation}
where the remainder $o(\rho ^{-m})$ has the property that
\begin{equation}
\lim_{\rho \rightarrow \infty} \rho \, o(\rho ^{-m}) = 0.
\end{equation}
A tensor field $T^{a_1...a_n}\,_{b_1...b_m}$ will be said to admit an asymptotic expansion to order 
$m$ if all its component in the \emph{Cartesian} chart $x^{a}$ do so. Its derivatives
$\partial_c T^{a_1...a_n}\,_{b_1...b_m}$ admit an expansion of order $m+1$.

\begin{figure}[h]
\begin{center}
 \includegraphics[width=8cm]{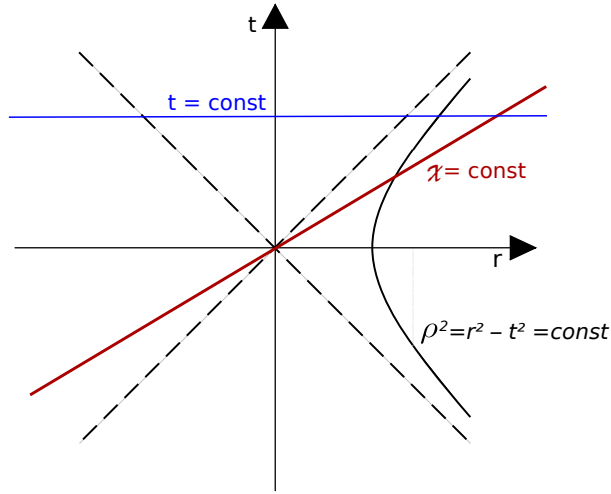}
  \caption{2D visualization of slices at constant $\chi$ and $t$ respectively.}\label{slices}
  \end{center}
\end{figure}

With these ingredients at hand we can now define an asymptotically flat spacetime in terms of its metric: 
a smooth spacetime metric $g$ on $\mathcal{R}$ is \emph{weakly asymptotically flat  at spatial 
infinity} if there exist a Minkowski metric $\eta$ such that outside a spatially compact world tube 
$(g - \eta)$ admits an asymptotic expansion to order 1 and $\lim_{\rho \rightarrow \infty } (g - \eta) = 0$.

In such a space-time the metric in the region $\mathcal{R}$ takes the form,
\begin{equation}
g_{ab} \d x^{a} \d x^{b} = \left(  1 +  \frac{2\sigma}{\rho} \right) \d \rho^{2} + 
2\rho\, \frac{\alpha_{i}}{\rho}\, \d \rho\, \d \Phi^{i} + \rho^{2} \left( h_{ij} + 
\frac{\,^{1} h_{ij}}{\rho} \right) \d \Phi^{i} \d \Phi^{j} + o(\rho^{-1})
\end{equation}
where $\sigma$, $\alpha_{i}$ and $\,^{1} h_{ij}$ only depend on the angles $\Phi^{i}$ and $h_{ij}$ is the metric 
on the unit time-like hyperboloid in Minkowski spacetime:
\begin{equation}\label{metric-hyperboloid}
h_{ij} \d \Phi^{i} \d \Phi^{j}  = -\d \chi^{2} + \cosh^2{\chi} (\d \theta^{2} + \sin^2{\theta} \d \phi^{2})\, .
\end{equation}

Note that also we could have expanded the metric in a chart $(r, \Phi)$, associated with a timelike 
cylinder, or any other chart. But we chose the chart $(\rho, \Phi)$ because it is well adapted to the 
geometry of the problem and will lead to several simplifications. In the case of a $3+1-$decomposition 
a cylindrical chart is a better choice (for details see \cite{C-JD}).

For this kind of spacetimes, one can always find another Minkowski metric such that its  
off-diagonal terms $\alpha_{i}$ vanish in leading order.  In \cite{aes} it is shown in detail that
the asymptotically flat metric can be written as
\begin{equation}\label{ds-AF}
\d s ^{2} = \left(  1 +  \frac{2\sigma}{\rho} \right) \d \rho^{2} +  \rho^{2}\, h_{ij} 
\left(  1 -  \frac{2\sigma}{\rho}  \right) \d \Phi^{i} \d \Phi^{j} + o(\rho^{-1}),
\end{equation}
with $\sigma (-\chi ,\pi -\theta ,\phi +\pi )=\sigma (\chi ,\theta ,\phi )$. We also see that $\,^{1} h_{ij} = 
-2 \sigma h_{ij}$. These two conditions restrict the asymptotic behaviour of the metric, but are necessary in order
to reduce the asymptotic symmetries to a Poincar\'e group, as demonstrated in \cite{aes}.

From the previous discussion and the form of the metric one can obtain the fall-off conditions for the tetrads.
As shown in \cite{aes} in order to have a well defined Lorentz angular momentum one needs to admit
an expansion of order 2. Therefore, we assume that in Cartesian coordinates we have the following behaviour
\begin{equation}\label{AFfalloff-tetrad}
e^{I} _{a} = \,^{o}e^{I} _{a} + \frac{\,^{1}e^{I} _{a}(\Phi )}{\rho} +  
\frac{\,^{2}e^{I} _{a}(\Phi )}{\rho^{2}} + o(\rho^{-2}),
\end{equation}
where $\,^{o}e^{I}$ is a fixed co-frame such that 
$g^{o}_{ab} = \eta_{IJ} \,^{o}e^{I} _{a} \,^{o}e^{I} _{b} $ is flat and $\partial_a(\,^{o}e^{I}_b)=0$. 

The sub-leading term $\,^{1}e^{I} _{a}$ can be obtained from (\ref{ds-AF}) and is given by \cite{aes},
\begin{equation}\label{1e}
\,^{1}e^{I} _{a} = \sigma(\Phi) (2 \rho_{a} \rho^{I} - \,^{o}e^{I} _{a} ) 
\end{equation}
where
\begin{equation}\label{rhoa-rhoI}
\rho_{a} = \partial_{a} \rho \,\,\,\,\,\,\,\,   \mathrm{and}  \,\,\,\,\,\,\,\,   \rho^{I} = \,^{o} e^{aI} \rho_{a}.
\end{equation}

The asymptotic expansion for connection can be obtained from the requirement that the connection be
compatible with the tetrad on $\mathcal{I}$, to appropriate leading order. This yields an asymptotic expansion 
of order 3 for the connection as,
\begin{equation}\label{AFfalloff-connection}
\omega_{a} ^{IJ} = \,^{o} \omega_{a} ^{IJ} + \frac{ \,^{1} \omega_{a} ^{IJ}}{\rho} + 
\frac{ \,^{2} \omega_{a} ^{IJ}}{\rho^{2}} + \frac{ \,^{3} \omega_{a} ^{IJ}}{\rho^{3}} + o(\rho^{-3})\, .
\end{equation}
We require that $De^I$ vanishes, to an appropriate order. More precisely, we ask that the term of order 0 in $De^I$
vanishes
\begin{equation}
\d \,^{o}e^{I} + \,^{o} {\omega^I}_K\w \,^{o}e^{K}=0\, ,
\end{equation}
and since $ \d \,^{o}e^{I}=0$ it follows that $\,^{o} \omega^{IK}=0$. The term of order 1 should
also vanish leading to $\,^{1} \omega^{IK}=0$. We also ask that the term of order 2 in $De^I$
vanishes, and we obtain
\begin{equation}\label{compatibility}
\d \bigl( \frac{\,^{1}e^{I}}{\rho}\bigr) =-\frac{ \,^{2} {\omega^I}_K}{\rho^{2}}\w \,^{o}e^{K}\, ,
\end{equation}
and we shall demand compatibility between $e$ and $\o$ only based on these conditions.
As a result, we obtain
\begin{equation}\label{2omega}
\,^{2} \omega_{a} ^{IJ} (\Phi) = 2 \rho^{2}\, \partial^{[J} (\rho^{-1} \,^{1}e^{I]} _{a} )
= 2 \rho\, ( 2 \rho^{[I} \rho_{a} \partial^{J]} \sigma - \,^{o}e^{[I} _{a} \partial^{J]} \sigma 
- \rho^{-1} \,^{o}e^{[I} _{a} \rho^{J]} \sigma )\, .
\end{equation}
Note that although $\rho$ appears explicitly in the previous expression, it is independent of $\rho$.
Therefore, in the asymptotic region we have $De^I = O(\rho^{-3})$.

\subsubsection{Palatini action with boundary term}

Now we have all necessary elements in order to prove the finiteness of the Palatini action with
boundary term, given by (\ref{PalatiniplusBoundary}). This expression can be re-written as,
\begin{equation}
S_{\mathrm{PB}}(e,\omega) = \frac{1}{2\kappa} \int_{\mathcal{M}} \left( \d \Sigma^{IJ} \wedge \omega_{IJ} - 
\Sigma^{IJ} \wedge \omega_{I}\,^{K} \wedge \omega_{KJ} \right)
\end{equation}
or in components,
\begin{equation}
S_{\mathrm{PB}}(e,\omega) = \frac{1}{4\kappa} \int_{\mathcal{M}} \left( \partial_{a} \Sigma_{bc}^{IJ} \omega_{dIJ} - 
\Sigma_{ab}^{IJ}  \omega_{cI}\,^{K} \omega_{dKJ} \right) \epsilon^{abcd}
\end{equation}
where  $\epsilon^{abcd}$ is the metric compatible 4-form on $\mathcal{M}$. The volume element 
is defined as $\epsilon^{abcd} = \tilde{\varepsilon}^{abcd} \,\, d^{4} x = -  \sqrt{|g |}\, \varepsilon^{abcd} \,\, \d^{4} x $, 
where $\tilde{\varepsilon}^{abcd}$ is the Levi-Civita tensor density of weight +1, while $\varepsilon^{abcd}$ is 
the Levi-Civita tensor\footnote{Note that $\varepsilon^{abcd} = s\, (\sqrt{|g|}\,)^{-1} \tilde{\varepsilon}^{abcd} $ with 
$s$ the signature of the metric, in our case $s=-1$.}.
We will prove that taking into account the boundary conditions (\ref{AFfalloff-tetrad}) and 
(\ref{AFfalloff-connection})
the action is manifestly finite always (even off-shell), if the two Cauchy surfaces are asymptotically time-translated 
with respect to each other.

Since $\sqrt{|g|} = e$, from the fall-off conditions on $e_{a} ^{I}$, it follows that asymptotically 
$e = \,^{0} e + O(\rho^{-1})$ where $\,^{0}e$ is the determinant of the fixed flat asymptotic metric, and since we 
are approaching the asymptotic region by a family of hyperboloids, it is natural to express it in hyperbolic coordinates. 
From (\ref{metric-hyperboloid}) and (\ref{ds-AF}), we obtain that 
$ \,^{0} e\, \d^4 x = \rho^{3} \cosh^2{\chi}\sin \theta\, \d \rho\, \d \chi\, \d \theta\, \d \phi 
=: \rho^{3} \, \d \rho\, \d^3\Phi $ and the volume element in the asymptotic region is of the form
$e\, \d^4 x=(1 + O(\rho^{-1}))\rho^3\, \d \rho\, \d^3\Phi$. It turns out that for our analysis it suffices to
take into  account only the leading term of the volume element.

In order to prove finiteness we shall consider the region bounded by two Cauchy 
slices, $M_1$ and $M_2$, corresponding to $t=\mathrm{const}$\footnote{We could have instead considered the 
region bounded by two surfaces corresponding to $\chi =\mathrm{const}$, 
but in that case for $\rho \rightarrow \infty$ the volume of the region does not need to converge (see Fig. \ref{slices}).}.
Since $t(\rho, \chi, \theta, \phi) = \rho \sinh \chi$ 
at the surface with constant $t$ we have $\rho\, \d \chi = -\tanh \chi \d \rho$. Substituting this into the metric 
 we can see that the leading term of the volume element of the asymptotic region of the Cauchy surface $M$ 
 is $\rho^{2} \sin \theta\, \d \rho\, \d \theta\, \d \phi$, since  
 as $\rho \rightarrow \infty $, the angle $\chi \rightarrow 0$.
 It follows that in the limit $\rho \rightarrow \infty$ the volume of the region $\mathcal{M}$ behaves as $\rho^2$. 

Now, we need to deduce the asymptotic behavior of
$\d \Sigma^{IJ} \wedge \omega_{IJ}=\epsilon_{IJKL}\d e^K\wedge e^L\wedge\omega^{IJ}$.
Since $\d (\,^{o}e^{I})=0$ it follows that 
\begin{equation}
\d e^K = \frac{1}{\rho}\, \d [\,^{1}e^{K} (\Phi )]+O(\rho^{-2})\, .
\end{equation}
The partial derivative, with respect to Cartesian coordinates, of any function  $f(\Phi )$ is proportional
to $\rho^{-1}$,
\begin{equation}
\partial_a  f(\Phi )=\frac{\partial\Phi^i}{\partial x^a}\,\frac{\partial f}{\partial\Phi^i}
= \frac{1}{\rho}\, A^i_a(\Phi )\frac{\partial f}{\partial\Phi^i}\, ,
\end{equation}
where the explicit expression for $A^i_a(\Phi )$ can be obtained from the relation between Cartesian
and hyperbolic coordinates. As a consequence $\d e^K = O(\rho^{-2})$, and since $\o_{IJ}=O(\rho^{-2})$
it follows that $\d \Sigma^{IJ} \wedge \omega_{IJ}$ falls off as $\rho^{-4}$, and the Palatini action
with boundary term is finite. 

Now let us prove the differentiability of the action (\ref{PalatiniplusBoundary}). 
As we have commented after (\ref{variationPalatini}), this action is differentiable if the boundary term
that appears in the variation vanishes. This boundary term is
\begin{equation}
\frac{1}{2 \kappa} \int_{\partial \mathcal{M}} \delta \Sigma^{IJ} \wedge \omega_{IJ} = 
\frac{1}{2 \kappa} \left( - \int_{M_{1}} + \int_{M_{2}} + \int_{\mathcal{I}} - 
\int_{\Delta} \right) \delta \Sigma^{IJ} \wedge \omega_{IJ}\, ,
\end{equation}
where we decomposed the boundary as $\partial \mathcal{M}=M_1\cup M_2\cup \mathcal{I}\cup\Delta$,
as in Fig.\ref{regionM}. 
On the Cauchy slices, $M_{1}$ and $M_{2}$, we assume 
$\delta e_{a}^{I} = 0$ so the integrals vanish, and in the following section we will prove that over 
$\Delta$ this integral also vanishes. Here we will focus on the contribution of the asymptotic region  
$\mathcal{I}$.

On a time-like hyperboloid $\mathcal{H}_\rho$, $\rho = \mathrm{const}$, the leading term of the volume element is 
$\rho^{3} \d^3\Phi$ and the boundary term can be written as,
\begin{equation}\label{Holst-I}
\frac{1}{2 \kappa}  \int_{\mathcal{I}} \delta \Sigma^{IJ} \wedge \omega_{IJ} 
=  - \frac{1}{4 \kappa}  \lim_{\rho\to\infty}{
\int_{\mathcal{H}_\rho} \delta \Sigma_{ab}^{IJ}\, \omega_{cIJ}\, \varepsilon^{abc} \rho^{3} \d^3\Phi}\, ,
\end{equation}
where $\varepsilon^{abc} = \rho_{d} \varepsilon^{dabc}$ is the Levi-Civita tensor on 
$\mathcal{H}_\rho$, with $\rho^{d}$ a unit normal to the surface $\rho = {\rm const}$.

Now we can use that,
\begin{equation}\label{deltaSigma-rho}
\delta \Sigma_{[ab]IJ} = \rho^{-1} \varepsilon_{IJKL} \,^{o} e_{[a|} ^{K} \delta \sigma (\Phi) 
\left(  2 \rho_{|b]} \rho^{L} - \,^{o} e_{|b]} ^{L} \right)+ O(\rho^{-2}) \, .
\end{equation}
Since, $\rho_{a} \varepsilon^{abc} =  \rho_{a} \rho_{d} \varepsilon^{dabc} = 0$, we obtain
\begin{equation}
 \delta \Sigma_{ab}^{IJ}\, \o_{cIJ}\, \varepsilon^{abc} =\frac{2}{\rho^3}\, \de\sigma\, \varepsilon_{IJKL} 
 \,^{o} e_{a}^{K} \,^{o} e_{b}^{L} \,^{o} e_{c}^{I} (\rho\, \partial^{J} \sigma +\rho^J\sigma )\varepsilon^{abc}
 + O(\rho^{-4}) \, .
\end{equation}
In this expression the term with a derivative of $\sigma$ is proportional to 
$\partial_{\rho} \sigma := \rho^{a} \partial_{a} \sigma = 0$, so that the 
variation (\ref{Holst-I}) reduces to 
\begin{equation}\label{final_variation}
 \frac{1}{2 \kappa}  \int_{\mathcal{I}} \delta \Sigma^{IJ} \wedge \omega_{IJ} =  -
 \frac{3}{2 \kappa} \de\, \left(\int_{\mathcal{H}_1} \sigma^2  \d^3\Phi \right)\, ,
\end{equation}
where $\mathcal{H}_1$ is the unit hyperboloid. So we see that the Palatini action with the boundary term
is differentiable when we restrict to configurations that satisfy asymptotically flat boundary conditions,
such that $C_\sigma :=\int_{\mathcal{H}_1} \sigma^2  \d^3\Phi$ has the same (arbitrary) value for all of them. 
In that case, the above expression (\ref{final_variation}) vanishes. This last condition is not an additional
restriction to the permissible configurations, because every one of them (compatible with our boundary
conditions) corresponds to some fixed value of $C_\sigma$. 

Here we want to emphasize the importance of the boundary term added to the action given that, without it, the action 
fails to be differentiable. The contribution from the asymptotic region to the variation of the \emph{Palatini action} is,
\begin{equation}
\frac{1}{2 \kappa}  \int_{\mathcal{I}}  \Sigma^{IJ}\wedge\delta\omega_{IJ} = \frac{1}{ 4 \kappa}  
\lim_{\rho\to\infty}\int_{\mathcal{H}_\rho} \Sigma_{ab}^{IJ}\,\delta \omega_{cIJ}\,\varepsilon^{abc} \rho^{3} \d^3\Phi\, .
\end{equation}
Our boundary conditions imply that $\Sigma_{ab}^{IJ} \delta \omega_{cIJ}=O(\rho^{-2})$, so that 
the integral behaves as $\int_{\mathcal{I}}  \rho\, \d^3\Phi$, and in the limit $\rho \rightarrow \infty$ 
is explicitly divergent.

\subsubsection{Holst and Nieh-Yan terms}\label{HolstAF}

As we have seen earlier, in the asymptotic region we have $De^I = O(\rho^{-3})$. Furthermore, as $D(De^I)=F^{IK}\w e_K$,
we have that $F^{IK}\w e_K = O(\rho^{-4})$. We can see that explicitly by calculating the term
of order 3 in this expression
\begin{equation}
F^{IK}\w e_K = \d \bigl( \frac{ \,^{2} {\omega^I}_K}{\rho^{2}}\bigr) \w \,^{o}e^{K}+ O(\rho^{-4})\, .
\end{equation}
The first term in the previous expression vanishes since $\d \bigl( \frac{ \,^{2} {\omega^I}_K}{\rho^{2}}\bigr) 
\w \,^{o}e^{K}=\d \bigl( \frac{ \,^{2} {\omega^I}_K}{\rho^{2}}\w \,^{o}e^{K} \bigr) = 0$, due to
(\ref{compatibility}). So, we see that the Holst term
\begin{equation}
S_{\mathrm{H}}= - \frac{1}{2 \kappa\g} \int_{\mathcal{M}} e^{I} \wedge e^{J} \wedge F_{IJ}\, ,
\end{equation}
is finite under these asymptotic conditions, since
$e^I\w e^J\w F_{IJ}$ goes as $\rho^{-4}$, while the volume element on every Cauchy surface goes as
$\rho^2\sin{\theta}\d\rho\,\d\theta\,\d\phi$.

The variation of the Holst term is well defined if the boundary term, obtained as a result of variation,
vanishes. We will analyze the contribution of this term
\begin{equation}\label{variation_boundary_Holst}
 \frac{1}{2\k\g}  \int_{\partial\mathcal{M}} e^I\w e^J \w \delta\o_{IJ}\, .
\end{equation}
Let us examine the contribution of the term of order 2 of the integrand in the integral over $\mathcal{I}$, it is
\begin{equation}\label{order2}
\,^{o}e^{I}\w \,^{o}e^{J}\w \frac{ \de (\,^{2} \o_{IJ})}{\rho^{2}}=\d\bigl[ \,^{o}e^{I}\w
\frac{\de (\,^{1}e_I)}{\rho}\bigr]\, ,
\end{equation}
due to (\ref{compatibility}) and $\d \,^{o}e^{I}=\de \,^{o}e^{I}=0$. The integral of (\ref{order2}) over $\mathcal{I}$
reduces to the integral of $\,^{o}e^{I}\w\frac{\de (\,^{1}e_I)}{\rho}$ over $\partial\mathcal{I}=S_{1\infty}\cup S_{2\infty}$,
where $S_{\infty}=\mathcal{I}\cap M$, and we see that this term does not contribute
to (\ref{variation_boundary_Holst}). So, the leading term in $e^I\w e^J \w \delta\o_{IJ}$ is of
order 3, and is proportional to
\begin{equation}
\,^{o}e^{I}\w \,^{1}e^{J}\w \de (\,^{2} \o_{IJ})+\,^{o}e^{I}\w \,^{o}e^{J}\w \de (\,^{3} \o_{IJ})\, .\notag
\end{equation}
Taking into account the expressions (\ref{1e}) and (\ref{2omega}) we can see that the first term vanishes, and the 
boundary term is of the form
\begin{equation}
\de S_{\mathrm{H}}\,\vert_\mathcal{I}=-\frac{1}{2\k}\,
\de\bigl( \int_{\mathcal{H}_1} \,^{o}e^{I}\w \,^{o}e^{J}\w \,^{3} \o_{IJ}\bigr)\, .
\end{equation}
This boundary term does not vanish (though it is finite), and it depends on $\,^{3} \o_{IJ}(\Phi )$, which is
not determined by our boundary conditions. Since we do not want to further restrict our asymptotic boundary conditions,
we should provide a boundary term for the Holst term, in order to make it
differentiable. As discussed in the subsection \ref{sec:4.2}, we have two possibilities, and we shall
analyze both of them. Let us consider first the boundary term $S_{\mathrm{BH}}$, given in (\ref{Holst-boundary-1}).
In order to show that this term is finite we should prove that the contribution  of order 2 vanishes. 
This contribution is given by
\begin{equation}
 \,^{o}e^{I}\w \,^{o}e^{J}\w \frac{\,^{2} \o_{IJ}}{\rho^{2}}=
 \d\bigl( \,^{o}e^{I}\w\frac{\,^{1}e_I}{\rho}\bigr)\, ,
\end{equation}
and due to the same arguments as in (\ref{order2}), we see that it does not contribute to the
boundary term (\ref{Holst-boundary-1}). So, the leading term of the integrand is of order 3,
and since the volume element at $\mathcal{H}_\rho$ goes as $\rho^3\d^3\Phi$, it follows that
(\ref{Holst-boundary-1}) is finite.

The Holst term with its boundary term (\ref{HolsttermBoundary}) can be written as
\begin{equation}
 S_{\mathrm{HB}}= - \frac{1}{2 \kappa\g} \int_{\mathcal{M}} e^{I} \wedge e^{J} \wedge F_{IJ}    
 +  \frac{1}{2 \kappa\g} \int _{\partial \mathcal{M}} e^{I} \wedge e^{J} \wedge \omega_{IJ}\, ,
\end{equation}
and also as an integral over $\mathcal{M}$
\begin{equation}
 S _{\mathrm{HB}}= - \frac{1}{2 \kappa\g} \int_{\mathcal{M}} 2\, \d e^{I} \wedge e^{J} \wedge \omega_{IJ}  
 -  e^{I} \wedge e^{J} \wedge \omega_{IK} \wedge \omega ^{K} \, _{J}\, .
\end{equation} 
As we have seen in (\ref{variationHolst}), the variation of the Holst term with its boundary term is
well defined provided that the following boundary contribution
\begin{equation}
\frac{1}{2 \kappa \gamma} \int_{\partial \mathcal{M}} \delta \Sigma^{IJ} \wedge \star\omega_{IJ} = 
\frac{1}{2 \kappa \gamma} \left( - \int_{M_{1}} + \int_{M_{2}} + \int_{\mathcal{I}} - 
\int_{\Delta} \right) \delta \Sigma^{IJ} \wedge \star \omega_{IJ}\, ,
\end{equation}
vanishes. We first note that $\omega^{IJ}_{a}$ 
and $\star \omega^{IJ}_{a}$ have the expansion of the same order, the leading term is $O(\rho^{-2})$.
Using (\ref{deltaSigma-rho}), the fact that $\rho_{a}$ is orthogonal to $\mathcal{I}$ and 
$\eta_{ab} \varepsilon^{abc} = 0$, one can see that the leading term in the integrand vanishes in
the asymptotic region,
so that $\delta \Sigma^{IJ} \wedge \star\omega_{IJ}=O(\rho^{-4})$ and the integral over
$\mathcal{I}$ vanishes.
In the next section we will prove that the integral over $\Delta$ vanishes, so that
$S_{\mathrm{HB}}$ is well defined.

The second choice for the well defined Holst term is $S_{\hny}$, given in (\ref{hny}). 
Let us first analyze the Neih-Yan topological term (\ref{Nieh-Yan}). It is easy to see that it
is finite since the integrand is of order 3, and the volume element on $\mathcal{H}$
is $\rho^3\d^3\Phi$, so the contribution at $\mathcal{I}$ is finite. 
The variation
of $S_{\ny}$ is
\begin{equation}
\de S_{\ny}=\int_{\partial\mathcal{M}}2De_I\w\de e^I-e^I\w e^J\w\de\o_{IJ}\, ,
\end{equation}
and we see that the first term vanishes, but the second one is exactly of the form that appears in (\ref{variation_boundary_Holst}),
and we have seen that it does not vanish, so the Nieh-Yan action is not differentiable.

As a result the combination of the Holst and Neih-Yan terms, $S_{\hny}$, is finite off-shell and its
variation is given by
\begin{equation}\label{var_torsion_term}
\de  S_{\hny} = -\frac{1}{\kappa\gamma} \int_{\mathcal{M}} \de e^I\w F_{IK}\w e^K + \de {\o^I}_K\w e^K\w De_I
 -\frac{1}{\kappa\gamma}\int_{\partial\mathcal{M}} De_I\w\de e^I\, .
\end{equation}
It is easy to see that this expression is well defined. Namely, the surface term vanishes since we demand $De^I=0$
on an isolated horizon $\Delta$, while at the spatial infinity the integrand behaves as $O(\rho^{-5})$ and
the volume element goes as $\rho^3\d^3\Phi$, and in the limit $\rho\to\infty$ the contribution of this
term vanishes. As a result, the action $S_{\hny}$ is well defined.

\subsubsection{Pontryagin and Euler terms}

Since we are interested in a generalization of the first order action of general relativity, that includes 
topological terms, we need to study their asymptotic behaviour. We will show that the Pontryagin and Euler terms are well defined.

It is straightforward to see that the Pontryagin term (\ref{Pontryagin}) is finite for asymptotically flat
boundary conditions. Since
\begin{equation}
S_{\Po} [e, \omega] = \frac{1}{4} \int_{\mathcal{M}} F_{ab}^{IJ} \wedge F_{cdIJ}\epsilon^{abcd}\, ,
\end{equation}
the finiteness of this expression depends on the asymptotic behavior of $F_{IJ}$.
Taking into account (\ref{AFfalloff-connection}), we can see 
that the leading term of $F_{abIJ}$ falls off as $\rho^{-3}$.
Since the volume of any Cauchy slice is proportional to $\rho^2$
in the limit when $\rho \rightarrow \infty$ the asymptotic contribution to the
integral goes to zero. As a result, the Pontryagin term is finite even off-shell. The same result holds for the Euler term 
(\ref{Euler}), since the leading term in the asymptotic form of $\star F_{IJ}$ is of the same order as
of $F_{IJ}$.

Now we want to prove that both terms are differentiable. As we have showed in (\ref{var_Po1}), 
the variation of the Pontryagin term is,
\begin{equation}
\de S_{\Po}=2 \int_{\partial \mathcal{M}}F^{IJ}\w\de\o_{IJ} = 2 \left( - \int_{M_{1}} + 
\int_{M_{2}} + \int_{\mathcal{I}} - \int_{\Delta} \right) F^{IJ} \wedge \delta \omega_{IJ}\, .
\end{equation}
In the following subsection we prove that on $\Delta$ the integral vanishes. For $\mathcal{I}$, 
we need to prove that the integral
\begin{equation}
\int_{ \mathcal{I}} F^{IJ} \w \de \o_{IJ} =- \lim_{\rho\to\infty}\int_{\mathcal{H}_\rho} 
F_{ab}^{IJ} \delta \omega_{cIJ} \varepsilon^{abc} \rho^{3} \d^3\Phi\, ,
\end{equation}
vanishes. Taking into account (\ref{AFfalloff-connection}) we can see that the leading term of 
$F_{abIJ} \omega_{c}\,^{IJ}$ goes as $\rho^{-5}$. 
Therefore the integral falls off as $\rho^{-2}$ which in the limit $\rho \rightarrow \infty$ goes to zero.
The same behavior holds for the Euler term, so it is also well defined.


\subsection{Internal boundary: Isolated horizons}
\label{sec:5.2}

We shall consider the contribution to the variation of the action at the internal
boundary, in this case a weakly isolated horizon. A weakly isolated horizon is
a non-expanding null 3-dimensional hypersurface, with an additional condition 
that implies that surface gravity is constant on a horizon. Let us specify with some
details its definition and basic properties \cite{afk, jaramillo}.

Let $\Delta$ be a 3-dimensional null surface of $(\mathcal{M},g_{ab})$, equipped 
with future directed
null normal $l$. Let $q_{ab}\,\hat{=}\, g_{\underleftarrow{ab}}$ be the (degenerate)
induced metric on $\D$ (we denote by $\hat{=}$ an equality which holds only on $\D$
and the arrow under a covariant index denotes the pullback of a corresponding form to
$\D$). A tensor $q^{ab}$ that satisfies $q^{ab}q_{ac}q_{bd}\,\heq\, q_{cd}$, is called an
inverse of $q_{ab}$. The expansion of a null normal $l$ is defined by $\theta_{(l)}=
q^{ab}\nabla_a l_b$, where $\nabla_a$ is a covariant derivative compatible with the
metric $g_{ab}$. 

The null hypersurface $\D$ is called a {\it non-expanding horizon} (NEH) if it satisfies the
following conditions: (i) $\D$ is topologically $S^2\times\re$, (ii) $\theta_{(l)}=0$
for any null normal $l$ and (iii) all equations of motion hold at $\D$  and 
$-T_{ab}l^b$ is future directed and causal for any $l$, where 
$T_{ab}$ is matter stress-energy tensor at $\D$. The second condition implies that
the area of the horizon is constant 'in time', so that the horizon is isolated. 

Let us analyze some properties of a NEH.
Since $l$ is a null normal to $\D$ its field lines are null geodesics. We define
surface gravity $\k_{(l)}$ as the acceleration of $l$
\be
l^a\nabla_a l^b\, \heq\, \k_{(l)}l^b\, .
\ee
We note that $\k_{(l)}$ is associated to a specific null normal $l$, if we replace
$l$ by $l'=fl$, where $f$ is an arbitrary positive function, the acceleration changes $\k_{(l')}=f\k_{(l)}+\La_l f$.
Also, since $l$ is normal to $\Delta$ its twist vanishes.
The condition $\theta_{(l)}=0$, together with the null Raychaudhuri and Einstein's equations and the condition on the 
stress-energy tensor imply that every $l$ is also shear free. Then, it follows that the horizon is `time' invariant,
in the sense that $\La_lq_{ab}\,\heq\, 0$.

As a basis for $T_p(\mathcal{M})$ it is convenient to use Newman-Penrose null-tetrad $(l,n,m,\bar{m})$, where
a null vector $n$ is transverse to $\Delta$, such that $l\cdot n=-1$, and
a complex vector field $m$ is tangential to $\D$, such that $m\cdot\bm =1$, and all the
other scalar products vanish. The pair $(m,\bar{m})$ forms a complex basis for $T_p(S_{\D})$, where
$S_{\D}$ is a compact two-dimensional cross section of $\Delta$. It can be shown that the area two-form on 
$S_{\D}$, defined as $\eps := im\w\bm$ is also preserved in `time', $\La_l\eps\,\heq\, 0$.

The geometry of a NEH is specified by $(q_{ab},\nabla_{\underleftarrow{a}})$, where $\nabla_{\underleftarrow{a}}$
is the unique connection induced from the connection in $\mathcal{M}$, $\nabla_a$, such that
$\nabla_{\underleftarrow{a}}q_{bc}=0$. Also, since the expansion, twist and shear of $l$ vanish,  
there exists a one-form $\o_a$, intrinsic to $\Delta$, defined as \cite{chandra, jaramillo}
\be
\nabla_{\ula{a}}l^b\, \heq\, \o_a l^b\, .\label{5.1}
\ee
Under a rescaling of the null normal $l\to l'=fl$, $\o$ transforms like a connection
$\o\to\o '=\o +\d\, ({\rm ln}f)$ (we see that $\o$ is invariant under constant rescaling).

We need one additional condition in order to satisfy the zeroth law of black hole
dynamics. Since $l$ can be rescaled by an arbitrary positive function, in general $\k_{(l)}$ is not 
constant on $\D$. At the other hand, it can be shown \cite{jaramillo} that  $\La_l\o_a\,\heq\, 
\nabla_{\underleftarrow{a}}\kappa_{(l)}$. 
If we want to establish the zeroth law of black hole dynamics
$\d\k_{(l)}\,\heq\, 0$ we need one additional condition, the `time' invariance of $\o$,
\begin{equation}
 \La_l\o\,\heq\, 0\, .
\end{equation}
Now, if we restrict to constant rescaling of $l$, $l\to l'=cl$ that leaves
$\o$ invariant, then the zeroth law of black hole dynamics follows,
for every null normal $l$ related to each other by constant rescaling.

All null normals related to each other by a constant rescaling form an equivalence class 
$[l]$. Now, we can define a {\it weakly isolated horizon} (WIH) $(\D ,[l])$ as
a non-expanding horizon equipped with an equivalence class $[l]$, such that
$\La_l\o\,\heq\, 0$, for all $l\in [l]$.

In order to analyze the contribution to the variation of the action over the internal 
boundary, which is a WIH $\D$, we equip $\D$ with a fixed class of null normals $[l]$
and fix an internal null tetrads $(l^I,n^I,m^I,\bm^I)$ on $\D$, such that their
derivative with respect to flat derivative operator $\partial_a$ vanishes.The
permissible histories at $\D$ should satisfy two conditions: (i) the vector field
$l^a:=e^a_Il^I$ should belong to the fixed equivalence class $[l]$ (this is a
condition on tetrads) and (ii) the tedrads and connection should be such that $(\D ,[l])$
constitute a WIH.

The expression for tetrads on $\D$ is given by 
\be
e_a^I\,\heq \, -l^I n_a + \bm^I m_a +m^I\bm_a\, ,\label{tetrad_horizon}
\ee
since $l_{\underleftarrow{a}}=0$. Using the relation $\epsilon^{IJKL}=i\, 4!\, l^{[I} n^J m^K \bm^{L]}$ we obtain the following
expression for the two-form $\Sigma^{IJ}$ 
\be
{\Sigma_{ab}}^{IJ}\,\heq\, 2l^{[I}n^{J]}\,\eps_{ab} +4i\, n_{[a} (m_{b]}\, l^{[I}\bm^{J]}-\bm_{b]}\, l^{[I}m^{J]})\, .\label{hor1}
\ee
The expression for the connection on $\D$ is given by \cite{cg}
\be
\o_{IJ}\,\heq\, -2\o\, l_{[I}n_{J]} + 2U\, l_{[I}\bm_{J]} + 2\bar{U}\, l_{[I}m_{J]}
+ 2V\, m_{[I}\bm_{J]}\, ,\label{hor2}
\ee
where we have introduced two new one-forms, a complex one $U$ and purely imaginary one $V$.
In \cite{cg} the expressions for these one forms is given in terms of Newman-Penrose (NP)
spin coefficients and null tetrads. First we have
\be
\o_a =-(\ve +\bar{\ve})n_a + (\bar{\a}+\b )\bm_a +(\a +\bar{\b})m_a\, ,
\ee
where $\a$, $\b$ and $\ve$ are NP spin coefficients. In what follows
we do not need their explicit form, for details see  \cite{cg}.
Since $\k_{(l)}=l^a\o_a$ it follows that $\k_{(l)}=\ve +\bar{\ve}$.

Also, it can be shown that \cite{afk}
\be\label{domega}
\d\o\,\heq\, 2\,{\rm Im}\, [\Psi_2]\,\eps\, ,
\ee
where $\Psi_2=C_{abcd}l^am^b\bm^cn^d$, and $C_{abcd}$ are the components of the Weyl tensor. 
Now, it is easy to see that the condition $\La_l\o\,\heq\, 0$ leads to $\d ({\rm Re}\, \ve )\,\heq\, 0$. 

On the other hand we have
\be
U_a\,\heq\, -\bar{\pi}n_a + \bar{\mu}m_a + \bar{\l}\bm_a\, ,
\ee
and
\be
V_a\,\heq\, -(\ve -\bar{\ve})n_a + (\b -\bar{\a})\bm_a +(\a -\bar{\b})m_a\, ,
\ee
where $\pi$, $\mu$ and $\l$ are additional NP spin coefficients. It has been also shown that
\be
\d V\,\heq\, {\mathcal F}\, \eps\, ,
\ee 
where ${\mathcal F}$ is the function of the Riemann curvature and Weyl tensor. Then, we
can calculate $\La_l V$,
\be
\La_l V=l\cdot\d V+\d (l\cdot V)=2\d ({\rm Im}\, \ve )\, ,\label{lv}
\ee
since $l\cdot\eps =0$.

We shall also need the expression for the pull-back of the curvature two-form on a NEH. The vanishing of
torsion leads to the relation between the curvature $F$ and the Riemann curvature $R$
\begin{equation}
{F_{ab}}^{IJ}={R_{ab}}^{cd}e_c^I e_d^J\, .
\end{equation}
Then, using the expression (\ref{tetrad_horizon}) and properties of the Reimann tensor on $\D$, one obtains that,
for a non-rotating NEH (the details are given in \cite{liko_booth})
\be
F^{IJ}\,\heq\, i\,\mathcal{R}\,\eps\, m^{[I}\,\bar{m}^{J]}+2F^{KL}n_L\, l^{[I} (m^{J]}m_K + \bar{m}^{J]}\bar{m}_K)\, ,
\label{curvature_horizon}
\ee
where $\mathcal{R}$ is the scalar curvature of $S_{\D}$.

Now we have all necessary elements in order to calculate the contribution of
the variation of the Palatini action, the Holst term and topological terms at WIH.
Before that, let us first examine the gauge invariance of the boundary terms of Palatini action and
Holst term, given in (\ref{PalatiniplusBoundary}) and (\ref{HolsttermBoundary}), on a
weakly isolated horizon $\D$. We have fixed the internal null basis $(l^I,n^I,m^I, \bm^I)$, but
we still have the residual Lorentz transformations, compatible with
the definition of $\D$, that act on tetrads and connection fields. The action of these transformations on the null
tetrad can be divided in two groups: ones that preserve the direction
of a vector $l$ and rotate $m$ 
\be
l^a\to c l^a\, ,\ \ \ n^a\to \frac{1}{c}n^a\, ,\ \ \ 
m^a\to e^{i\theta}m^a\, ,\label{lorentz1}
\ee
and ones that leave $l$ invariant, but change $n$ and $m$
\be
l^a\to l^a\, ,\ \ \ n^a\to n^a-u m^a-\bar{u}\bm^a
+u\bar{u}l^a\, ,\ \ \ m^a\to m^a-\bar{u} l^a\, .\label{lorentz2}
\ee
Note that $c=\mathrm{const}>0$, since $l\in [l]$, while $\theta$ and $u$ are arbitrary real and complex functions, respectively. 
We shall show later that the NP coefficient $\varepsilon$
transforms under (\ref{lorentz1}) so that by choosing 
$\theta$  such that $l^a\nabla_a\theta =-2\, ({\rm Im}\ve )$ the new $\tilde{\varepsilon}$ becomes real. 
This restricts the remaining gauge freedom 
$m\to e^{i\tilde{\theta}}m$ to the functions $\tilde{\theta}$ of the form  $\nabla_a\tilde{\theta}\,\heq\,w m_a+
\bar{w}\bm_a$, where $w$ is arbitrary. Note also that there are no restrictions on $u$ in (\ref{lorentz2}).

The Palatini boundary term on the horizon reduces to
\be
\frac{1}{2\k}\int_\D\S_{IJ}\w\o^{IJ}=\frac{1}{\k}\int_\D\eps\w\o\, .
\ee
It was shown in \cite{bcg} that $\o$ is invariant under both classes of transformations.
Two-form $\eps =im\w\bm$ is invariant under (\ref{lorentz1}), and also under (\ref{lorentz2}) 
since this transformation implies  $m_a\to m_a$, due to $l_{\underleftarrow{a}}\,\heq\, 0$. 

Similarly, the Holst boundary term on the horizon is
\be
\frac{1}{2\k\g}\int_\D\S_{IJ}\w\star\o^{IJ}=\frac{i}{\k}\int_\D\eps\w V\, .
\ee
It turns out that $V$ is invariant under (\ref{lorentz2}), and under (\ref{lorentz1}) it transforms
as $V\to V-i\d\tilde{\theta}$. Since $\nabla_a\tilde{\theta}\,\heq\,w m_a+
\bar{w}\bm_a$, we have $\eps\w\d\theta\,\heq\, 0$. As a consequence $\eps\w V$ 
is also invariant under gauge transformations on $\D$.

\subsubsection{Palatini action and isolated horizons}

In this part we will analyze the variation of the Palatini action with boundary term (\ref{PalatiniplusBoundary}), 
on an isolated horizon $\D$
\be
\de S_{\mathrm{PB}}\vert_\D = \frac{1}{2\k}
\int_{\D} \epsilon_{IJKL}\o^{IJ}\w e^K\w\de e^L =
\frac{1}{2\k}\int_{\D} \o \w\de\,\eps\, .\label{var1}
\ee
Since $\D$ is a non-expanding horizon, $\La_l\,\eps\,\heq\, 0$. 
Any other permissible configuration of tetrads, $(e_I^a)'$, should
also satisfy $\La_{l'}\eps '\,\heq\, 0$, where $l'^a\in [l]$ and $\eps '=\eps +\de\,\eps$.
For the null normals in the equivalence class $[l]$, $\La_{l'}\eps '=c\,\La_l\eps '\,\heq\, 0$,
and it follows that $\La_l\,\de\,\eps\,\heq\, 0$. In the variational principle all fields are fixed 
on initial and final Cauchy surfaces, $M_1$ and $M_2$, in particular $\de\,\eps =0$ on two-spheres at
the intersection of the initial and final Cauchy surface with the WIH, 
$S_{1,2} := M_{1,2}\cap \D$ (see Fig. 1). Furthermore, $\de\eps$ does not change along any null normal $l$, so that 
$\de\,\eps\,\heq\, 0$ on the entire horizon (comprised between the two Cauchy surfaces) and the integral 
(\ref{var1}) vanishes. We should remark that, in the following parts, we {\it will} use the same argument whenever we 
have some field configuration whose Lie derivative along $l$ 
vanishes on the horizon, to prove that its variation is zero on the horizon.

We note that the variation of the Palatini action, without boundary term, at $\D$ is
\be
\de S_{\mathrm{P}}\vert_{\D} =-\frac{1}{2\k} \int_{\D} \epsilon_{IJKL}\de\o^{IJ}\w e^K\w e^L =
-\frac{1}{2\k}\int_{\D} \de\o  \w\eps\, .\label{var1b}
\ee
In this case, one can argue that the term on the RHS vanishes, because from $\La_l\o\,\heq\, 0$ it follows
that $\de\o\,\heq\, 0$ (a similar, but slightly different, argument was used in \cite{afk}). 
We see that the variational
principle for the Palatini action is well defined even without boundary terms on the horizon. 
Nevertheless, for the reasons already mentioned in the previous section we shall keep the boundary terms in (\ref{PalatiniplusBoundary})
on the whole boundary, including the internal one.

\subsubsection{Holst term and isolated horizons}

Let us now analyze the variation of the Holst term on a horizon. We have
\be
\de S_{\mathrm{H}}\vert_\D = -\frac{1}{2\k\g}\int_{\D} \de\o^{IJ}\w e_I\w e_J =
-\frac{i}{2\k\g}\int_{\D} \de V \w\eps\, .\label{var2b}
\ee
In this case we can not use the same argument as in the case of Palatini action
since the Lie derivative of $V$ does not 
vanish on $\D$, as shown in (\ref{lv}), $\La_l V\,\heq\, 2\,\d\, ({\rm Im}\ve )$.  
As we commented earlier, we have a freedom to perform local Lorentz transformations 
in order to make $\ve$ a real function. Namely, the rotation in the $(m,\bm )$ plane, given
by $m\to e^{i\theta}m$, where $\theta$ is an arbitrary function, generates the following transformation
of the NP spin coefficient $\ve$ \cite{chandra}: $ \ve\to\ve +\f {i}{2}l^a\nabla_a\theta$.
So, $\ve$ can be made real after the appropriate rotation that satisfies the 
condition $l^a\nabla_a\theta =-2\, ({\rm Im}\ve )$. Due to this gauge freedom we can always
choose a real $\ve$, and as a result $\La_l V=0$. When we change the configuration of fields this
condition could be violated, namely the new $\ve'$ need not be real. Only if we restrict the variations, by
demanding $\de ({\rm Im}\ve )= {\rm const.}$, we obtain
$\La_{l'}V'=0$. Then, using the same arguments as before we can conclude that $\de V\,\heq\, 0$, in
the variational principle and (\ref{var2b}) also vanishes. 
As a result we see that the Holst term by itself is not differentiable on an isolated horizon, for arbitrary 
allowed variations of fields, but we will show that by adding an appropriate boundary term it can be made well 
defined. For a more detailed discussion see \cite{CReV}.

Let us now consider the contributions coming from the Holst term, with its boundary term, given in (\ref{HolsttermBoundary}), 
on an isolated horizon
\be
\de S_{\mathrm{HB}}\vert_\D = \frac{1}{2\k\g}
\int_{\D} \o^{IJ}\w e_I\w\de e_J =
\frac{i}{2\k\g}\int_{\D} V\w\de\,\eps\, ,\label{var2}
\ee
and for the same reasons that we used before, after the equation (\ref{var1}), since $\La_l\,\eps\,\heq\, 0$, 
it follows that $\de\,\eps\,\heq\, 0$ on the horizon $\Delta$, and the variation (\ref{var2}) 
vanishes. 

At the end, the variation of the other choice for a well defined Holst action, which is quadratic in torsion, 
$S_{\hny}$, on the isolated horizon vanishes since
\begin{equation}
\de S_{\hny}\vert_\D = -\frac{1}{\k\g} \int_{\D} De_I\w\de e^I = 0\, ,
\end{equation}
since $De^I\,\heq\,0$.

\subsubsection{Pontryagin and Euler terms and isolated horizon}

Let us now consider the possible contributions coming from the remaining topological terms. That is,
we shall see whether the above conditions are sufficient to make their variation well defined at $\D$. 
The variation of the Pontryagin term on the non-rotating horizon is 
\be
\de S_{\mathrm{Po}}\vert_\D = 2\int_{\D}F^{IJ}\w\de\o_{IJ}
=-2i\int_{\D}\mathcal{R}\, \eps\,\w\de V\, ,\label{var3}
\ee
where we have used the expresions for the curvature at the horizon (\ref{curvature_horizon}) and
for the connection (\ref{hor2}). 
The argument just presented in the previous 
part implies that $\delta V\, \hat{=}\, 0$, for the variations that satisfy  $\delta ({\rm Im}\varepsilon )= {\rm const.}$, 
so, under this condition, the variation $\delta S_{\mathrm{Po}}$ vanishes at the horizon.

The variation of the Euler term on the non-rotating WIH is 
\be
\de S_{\mathrm{E}}\vert_\D =
2\int_{\D}\star\, F^{IJ}\w\de\o_{IJ}= 2\int_{\D} \mathcal{R}\,\eps\,\w\de\o\, ,\label{var4}
\ee
where $\star\, F^{IJ}:=\frac{1}{2}{\epsilon^{IJ}}_{KL}F^{KL}$, so from (\ref{curvature_horizon}) it follows
\begin{equation}
\star F^{IJ}\,\heq\, \mathcal{R}\,\eps\, l^{[I}\, n^{J]}+2i\, F_{MN}n^N\, l^{[I} (\bm^{J]}m^M + m^{J]}\bar{m}^M)\, .
\label{dual_curvature_horizon} 
\end{equation}
The variation of the Euler term vanishes since $\de\o\,\heq\, 0$.

We can then conclude that the inclusion of the topological terms to the action is compatible with a well 
defined action principle, without the need of adding new boundary terms at the horizon.

\subsection{The complete action}
\label{sec:4.4}

In this section we have introduced boundary conditions for the gravitational field at infinity and at an internal boundary, and have
analyzed the contribution from the different terms that one can add to the action.

We saw that both the Palatini and Holst terms need to be supplemented with a boundary term to make them finite and differentiable.
As we saw, both Pontryagin and Euler terms, 
$S_{\mathrm{Po}}$ and $S_{\mathrm{E}}$ respectively, are well defined for our boundary conditions.
 This means that we can add them to 
$S_{\rm{PB}}$, and the resulting action will be again well defined. 
The complete action is,
\begin{equation}\label{complete-action2}
S[e, \omega] = S_{\rm PB} + \alpha_{1} S_{\rm Po} + 
\alpha_{2}S_{\rm E} + S_{\rm H} + \alpha_{3} S_{\rm NY} + \alpha_{4}S_{\rm BH}\, .
\end{equation}
Let us now comment on the restrictions on the coupling constants that arose from our previous considerations.
The coupling constants $\alpha_{1}$ and $\alpha_{2}$, are not fixed by our boundary conditions,
while different choices for the Holst-Nieh-Yan sector of the theory, discussed in the previous part, 
imply particular combinations of $\alpha_3$ and $\alpha_4$. To see that, consider the term
$S_{\rm BH}$ given in (\ref{Holst-boundary-1}). 
As we have seen in the previous part, if we choose  $\a_3=-\frac{1}{2\k\g}$ then the combination of the Holst
and Nieh-Yan terms gives $S_{\hny}$, which is well defined and no additional boundary term is needed, so in that case 
we have $\a_4=0$. For every other value of $\a_3$ we need to add a boundary term $S_{\rm BH}$, and in that case 
we obtain that $\a_4=\frac{1}{2\k\g}+\a_3$.
Apart from these special cases, there is no other non-trivial relation between the different coupling constants.

This feature has to be contrasted with other asymptotic conditions studied in the literature, in particular the AdS asymptotic conditions (that we shall, 
however, not consider here).
It turns out that the Palatini action with a negative cosmological constant term ${\Lambda}$, is not well defined for 
asymptotically anti-de Sitter (AAdS) spacetimes, but it can be made differentiable after the 
addition of an appropriate  boundary term. Several different proposals and approaches have appeared
in the literature. Here she shall briefly mention some of them without attempting to be exhaustive. 
In \cite{apv}
it is shown that one can make the action well defined by adding the same boundary term as in the asymptotically flat case, with an appropriately modified coupling constant. 
Second, there have been proposals to link topological terms to
the AdS/CFT correspondence. As shown in \cite{aros} and \cite{rodrigo}, one can choose the Euler 
topological term as a boundary term in order to make the action differentiable, and that choice fixes the value of $\alpha_2$. 
In that case $\alpha_2\sim \frac{1}{\Lambda}$, and the asymptotically flat case cannot be obtained in the limit $\Lambda\to 0$. 
The Pontryagin term can also appear naturally when self-duality for the Weyl tensor is used as a boundary condition \cite{selfdual} 
(see also \cite{mansi}).
This provides a topological interpretation to the holographic stress tensor/Cotton tensor introduced in \cite{haro} and \cite{bakas}. 
This condition fixes the value of $\alpha_1$ and it also  turns out to be
inversely proportional to the cosmological constant. Again, the asymptotically flat case can not be recovered as a limiting case. 
This provides yet another clue that the AAdS and asymptotically flat cases  are not connected via a limiting procedure. A possible explanation
for this fact is that, in the AF case, the fall-off conditions for the fields make both topological terms decay fast enough
so that they are both well defined and finite, whereas in the AAdS case that does not seem to be the case.
Certainly, a more detailed study of this relation is called for. 
Finally, the differentiability of  the Nieh-Yan term in the case of AAdS spacetimes has been analyzed in \cite{sengupta}. 
The result is that it becomes well defined only after the addition of the Pontryagin term, 
with an appropriate coupling constant. 
It is important to remark that the details of the asymptotic behaviour in \cite{apv} are different from those in the other mentioned papers, 
so one must proceed with care when comparing all the results here mentioned.

\section{Conserved quantities: Hamiltonian and Noether charges}
\label{sec:6}

So far we have analyzed the action principle for gravity in  the first order formulation with
two possible boundary conditions. We have seen that the action principle is well defined for 
a suitable choice of boundary terms even in the case when topological terms are 
incorporated. In this part we shall extract some of the information that comes from the 
covariant Hamiltonian formulation,  such as the associated conserved quantities. 
As we have discussed in Secs.~\ref{sec:3.2} and \ref{sec:3.3} there are 
two classes of quantities, namely those that are generators of Hamiltonian symmetries 
and the so called Noether charges. It is illustrative to
analyze the relation between the Hamiltonian and Noether charges
for the most general first order gravitational action, focusing on the role that the boundary terms play.
As could be expected, the fact that the boundary terms do not contribute to the symplectic structure implies 
that the Hamiltonian charges are insensitive to the existence of extra 
boundary terms. However, as we review in detail in what follows, the Noetherian quantities {\it do} depend on 
the boundary terms. Specifically, we are interested
in the relation of the Noether charge with the energy at the asymptotic region and the
energy of the horizon. Here, we shall recall the principal results of this analysis (more details can be found
in \cite{crv1}).

\subsection{Symplectic structure and energy}
\label{sec:6.1}

Let us start with the Hamiltonian charge related to diffeomorphisms, in the first order formalism,
for asymptotically flat configurations. The first step is the construction of the pre-symplectic structure $\bar{\Omega}$
for the full theory, given by (\ref{complete-action}).
As we have seen in (\ref{defJ}) the boundary terms in the action (topological terms) do not contribute to the symplectic current $J$, 
so that the only contributions in our case come 
from Palatini action and the Holst term\footnote{Note that there have been some
statements in the literature claiming that topological terms contribute to the symplectic structure when there are 
boundaries present \cite{liko_2, perez_2}.}. For this reason we shall only consider the Palatini and Holst terms in this part.
From the equation (\ref{intJzero}) one can obtain a conserved pre-symplectic structure, as an integral of $J$ over a spatial surface, if the integral
of the symplectic current over the asymptotic region vanishes and if the integral over an isolated horizon
behaves appropriately. 

As shown in \cite{aes}, for Palatini action and asymptotically flat spacetimes, $\int_{\mathcal{I}}J_P=0$, where $J_P$ is
given by (\ref{JPalatini}). On the horizon we have
\begin{equation}
J_{\mathrm{P}}(\delta_{1}, \delta_{2})\, \vert_\Delta  = -\frac{1}{\kappa}\, \delta_{[1} \Sigma^{IJ} \wedge \delta_{2]} \omega_{IJ} 
\, |_\Delta = -\frac{2}{\kappa}\, \de_{[1}(\eps )\w\de_{2]}\o \, ,
\end{equation}
where we have used the expressions (\ref{hor1}) and (\ref{hor2}), for $\Sigma^{IJ}$ and $\o_{IJ}$ on $\D$. 
Now, for a non-expanding horizon $\d\,\eps =0$. Let us now derive the horizon contribution to the symplectic structure for the case 
of a non rotating horizon. In this case, $\d\,\o =0$, so we can define 
a potential $\psi$, such that $\o =\d\,\psi$. As a result, we obtain
\begin{equation}
\de_{1}(\eps )\w\de_{2}\o =  \d\, (\de_{1}(\eps )\,\de_{2}\psi )\, ,
\end{equation}
so that
\be
\int_\D J_P(\de_1,\de_2)=\frac{1}{\k}\int_{\partial\D}\de_1\psi\,\de_2 (\eps )-\de_2\psi\,\de_1 (\eps )\, .\label{JPalhorizon}
\ee
Note that $\psi$ is defined up to a constant that we shall now fix. First, note that $\La_l\psi = l\cdot\d\psi=l\cdot\o=\k_{(l)}$ 
(we also take this identity as a definition for the scalar $\psi$  in the general, rotating case). 
Furthermore, with no loss of generality, we can fix the arbitrary constant in $\psi$ such that
 $\psi =0\ \ {\rm on}\ \ S_{1\D}$, 
with $S_{1\D}=M_1\cap\D$, 2-sphere at the intersection of a Cauchy surface with a horizon. 
For a rotating horizon one should note that the same expression 
(\ref{JPalhorizon}) for the symplectic structure arises, as shown in  \cite{afk}. Note also that the integral 
(\ref{JPalhorizon}) reduces to a surface integral over $\partial\D =S_{1\D}\cup S_{2\D}$. 

Finally, as explained in the subsection \ref{sec:3.1}, the general form of the pre-symplectic structure  given in 
(\ref{sympl_struct}), is the sum of two integrals, one of them over a Cauchy surface and the other one over 
a 2-sphere at the internal boundary, which in this case  is a section of an isolated horizon. 
Taking into account the expression for the symplectic current (\ref{JPalatini}), 
and the result given above in (\ref{JPalhorizon}), the pre-symplectic structure for the Palatini action,
for asymptotically flat spacetimes with weakly isolated horizon, takes the form \cite{afk}
\begin{equation}
\bar{\O}_P(\de_1,\de_2)=-\frac{1}{2\k}\int_{M}\de_1\S^{IJ}\w\de_2\o_{IJ} - \de_2\S^{IJ}\w\de_1\o_{IJ} 
-\frac{1}{\k}\int_{S_\D}\de_1\psi\,\de_2 (\eps )-\de_2\psi\,\de_1 (\eps )\, .\label{pal_ss}
\end{equation}
We see that the existence of an isolated horizon modifies the symplectic structure of the theory. 

We have seen in previous sections that the symplectic current of the Holst term is a total derivative given by (\ref{symplectic_current_holst}).
As we have seen in the Sec.~\ref{sec:3}, when the symplectic current is a total derivative, the covariant
Hamiltonian formalism indicates that the corresponding (pre)-symplectic structure vanishes.
As we also remarked in Sec.~\ref{sec:3}, one could postulate a conserved two form $\tilde\O$
if $\int_{\mathcal I}J_{\mathrm{H}}=0$ and $\int_{\D}J_{\mathrm{H}}=0$, in which case
this term defines a conserved pre-symplectic structure. Let us, for completeness, consider this
possibility. In \cite{cwe} it has been shown that the integral at ${\mathcal I}$
vanishes, so here we shall focus on the integral over $\Delta$
\begin{equation}\label{Holst_int_horizon}
 \int_{\D}J_{\mathrm{H}}=\frac{1}{\k\g}\int_{\partial\D}\de_1 e^I\w\de_2 e_I=\frac{1}{\k\g}\int_{\partial\D}
 \de_1 m\w \de_2\bm + \de_1\bm\w\de_2 m\, .
\end{equation}
We can perform an appropriate Lorentz transformation at the horizon in order to get a foliation of $\D$ spanned
by $m$ and $\bm$, that is Lie dragged along $l$ \cite{afk}, that implies $\La_l m^a\,\heq\, 0$. On  the other hand,
$\partial\D =S_{\Delta 1}\cup S_{\Delta 2}$, so it is sufficient to show that the integrand in (\ref{Holst_int_horizon})
is Lie dragged along $l$. The variations in (\ref{Holst_int_horizon}) are tangential to $S_{\D}$, hence we have
$\La_l\de_1 m = \de_1\La_l m=0$, so that the integrals over $S_{\Delta 1}$ and $S_{\Delta 2}$ are equal and
$\int_{\D}J_{\mathrm{HB}}=0$. So we can define a conserved pre-symplectic structure corresponding to the Holst term
\begin{equation}
 \tilde{\O}_H(\de_1,\de_2)=\frac{1}{\k\g}\int_{\partial M}\de_1 e^I\w\de_2 e_I\, ,
\end{equation}
where the integration is performed over $\partial M=S_{\infty}\cup S_{\D}$. As shown in \cite{cwe}, the integral over
$S_{\infty}$ vanishes, due to asymptotic conditions, and the only contribution comes from $S_{\D}$. Finally, we see that the quantity
\begin{equation}\label{symplectic_structure_holst}
 \tilde{\O}_H(\de_1,\de_2)=\frac{1}{\k\g}\int_{S_{\D}}\de_1 e^I\w\de_2 e_I\, .
\end{equation}
defines a conserved two-form. Note that this is precisely 
the symplectic structure for the Holst term defined in \cite{merced-holst}, even though  it was not explicitly shown
that it is independent of $M$ (this result depends on the details of the boundary conditions).\footnote{Recall that $\tilde{\O}$ 
does not follow from the systematic derivation of the covariant Hamiltonian formalism of Sec.~\ref{sec:3}. Also, one should remember
that in Sec.~\ref{sec:4}
we showed that for the Pontryagin and Euler terms, the introduction of a non-trivial $\tilde{\O}$ leads to inconsistencies, so one 
should be careful when postulating such object.}

Let us now recall the construction of the conserved charges for this theory.
We shall consider the Hamiltonian $H_\xi$ that is a conserved quantity corresponding
to  asymptotic symmetries and symmetries on the horizon of a spacetime. Our asymptotic conditions are
chosen in such a way that the asymptotic symmetry group be the Poincar\'e group. The corresponding
conserved quantities for the Palatini action, namely energy-momentum and relativistic angular momentum, are constructed in \cite{aes}.
The contribution to the energy from a weakly isolated horizon has been analyzed in \cite{afk}, where the 
first law of mechanics of non-rotating black holes was first deduced. Rotating isolated horizons are
studied in \cite{abl}, where the contribution from the angular momentum of a horizon has
been included. In this paper we restrict our attention to energy and give a review of the principal results presented in \cite{afk}.

Let us consider a case when $\xi$ is the infinitesimal generator of asymptotic time 
translations of the spacetime. It induces time evolution
on the covariant phase space, generated by a vector field $\de_\xi :=(\La_\xi e,\La_\xi\o )$.
At infinity $\xi$ should approach a
time-translation Killing vector field of the asymptotically flat spacetime. On the 
other hand, if we have a non-rotating horizon $\D$, then $\xi$, at the horizon, should belong to the 
equivalence class $[l]$. In order that $\de_\xi$ represents a phase space symmetry the condition
$\La_{\de_\xi}\bar{\O} =0$ should be satisfied. As we have seen in Sec.~\ref{sec:3.2}, $\de_\xi$ is a Hamiltonian vector field iff the one-form
\be
X_\xi (\de )=\bar{\O} (\de ,\de_\xi )\, ,
\ee
is closed, and the Hamiltonian $H_\xi$ is defined as
\be
X_\xi (\de )=\de H_\xi\, .
\ee
In the presence of an isolated horizon, the symplectic structure for the Palatini action (\ref{pal_ss}) has two contributions, 
one from the Cauchy surface $M$ and the 
other one from the two-sphere $S_\D$, at the intersection of $M$ with $\D$. 
Nevertheless,  the integral over $S_\D$ in $\bar{\O}_P (\de ,\de_\xi )$ vanishes, as shown in \cite{crv1,afk}.
As we have repeatedly stressed, the contribution to the symplectic structure from the Holst and topological terms is
trivial, so it is enough to consider the expression coming from the Palatini action\footnote{However, if one were to
postulate a contribution to the symplectic structure $\tilde{\O}_H$ coming from the Holst term (\ref{symplectic_structure_holst}) 
one would obtain  that $\tilde{\O}_H (\de ,\de_\xi )=0$ \cite{crv1}.}.
As a result $\de H_\xi := \bar{\O} (\de ,\de_\xi )=\bar{\O}_P(\de ,\de_\xi )$ only has a contribution from the Palatini action.
It turns out that the integrand in $\bar{\O}_P (\de ,\de_\xi )$ is a total derivative \cite{afk}, so that
\be
\de H_\xi 
=-\frac{1}{2\k}\int_{\partial M}(\xi\cdot\o^{IJ} )\de\S_{IJ} -
(\xi\cdot\S_{IJ} )\w\de\o^{IJ} \, ,\label{firstlaw}
\ee
where the integration is over the boundaries of the Cauchy surface $M$,
the two-spheres $S_\infty$ and $S_\D$. 

The integral at infinity vanishes for every permissible
variation $\de$, if and only if $\xi$ vanishes asymptotically, so that only diffeomorphisms which preserve
the boundary conditions and which are identity at infinity are gauge transformations, i.e. they
are in the kernel of $\bar{\O}$. The asymptotic
symmetry group is the quotient of the group of space-time diffeomorphisms which preserve the 
boundary conditions by its subgroup consisting of asymptotically identity diffeomorphisms. In the asymptotically flat case 
this is the Poincaré group and its action generates canonical transformations on the covariant
phase space whose generating function is denoted by $H_\xi^\infty$. The situation is similar at the horizon $\D$
and infinitesimal diffeomorphisms need not be in the kernel of the symplectic structure unless they vanish
on $\D$ and the horizon symmetry
group is the quotient of the Lie group of all infinitesimal space-time diffeomorphisms
which preserve the horizon structure by its subgroup consisting of elements which are identity
on the horizon \cite{abl}. The corresponding generating function is denoted by $H_\xi^\Delta$.

The surface term at infinity in the expression (\ref{firstlaw}) defines the gravitational energy at
the asymptotic region, whose variation  is given by the expression
\begin{equation}\label{energy_infinity}
\de E_\xi^\infty = 
\frac{1}{2\k}\int_{S_\infty}(\xi\cdot\S_{IJ} )\w\de\o^{IJ}\, ,
\end{equation}
where the first term in the above expression vanishes
due to the asymptotic behaviour of the tetrad and connection.
As shown in \cite{aes}, this expression represents
the variation of the ADM energy, $\de E_{\xi\mathrm{ADM}}$, associated with the asymptotic 
time-translation defined by $\xi$. Thus,
\begin{equation}
 E_\xi^\infty = E_{\xi\mathrm{ADM}}=\frac{2}{\k}\int_{S_\infty}\sigma\,\d^2 S_o\, ,
\end{equation}
where $\d^2 S_o$ is the area element of the unit 2-sphere. 

Now, the surface term at the horizon in the expression (\ref{firstlaw}) defines the  
horizon energy associated to  the time translation $\xi$, whose variation is given by
\begin{equation}\label{energy_horizon}
\de E_\xi^\D  
= \frac{1}{2\k}\int_{S_\D}(\xi\cdot\o^{IJ} )\,\de\S_{IJ}\, ,
\end{equation}
since the second term in (\ref{firstlaw}) vanishes at the horizon. The remaining term, when computed on the horizon is
of the form
\begin{equation}
\de E_\xi^\D = \frac{1}{\k}\,\int_{S_{\D}}(\xi\cdot\o )\,\de(\eps )= \frac{1}{\k}\,\k_{(\xi )}\,\de a_\D
\, ,
\end{equation}
since $\xi\cdot\o =c\, l\cdot\o =c\k_{(l)}=\k_{(\xi )}$ is constant on the horizon and where  $a_\D =\int_{S_{\D}}\eps$ is the area 
of the horizon. 

We can now see that the expression 
 (\ref{firstlaw}) encodes the first law of mechanics for non-rotating black holes, since
it follows that
\be\label{first law}
\de H_\xi =\de E_{\xi\mathrm{ADM}}-\frac{1}{\k}\,\k_{(\xi )}\de a_\D\, .
\ee
It is important to note that the necessary condition for the existence
of $H_\xi$ is that the surface gravity, $\k_{(\xi )}$, be a function only of the horizon area $a_\D$.
In that case
\be
H_\xi =E_{\xi\mathrm{ADM}}-E_\xi^\D\, .
\ee

In the following section we shall recall the construction of the Noether charge that corresponds to time translation,
for each term of the action (\ref{complete-action}). Contrary to the Hamiltonian conserved quantities, the contribution 
from the topological terms might actually be non-vanishing. 

\subsection{Noether charges}

In this section we shall review the main results of the 
study of the Noether charges for the most general first order action we have been considering. We shall follow \cite{crv1} closely, where details can be found. 
It has two parts. In the first one, we focus our attention on the Palatini action, while in the second part 
we look at the Holst and topological terms.

Let us start by recalling the relation between Hamiltonian and Noether charges given by the covariant formalism. 
We have just seen that $\de H_\xi$ is
an integral over a Cauchy surface of the symplectic current $J(\de ,\de_\xi )$.
In section~\ref{sec:3.3} we displayed the relation between the symplectic and Noether currents,
given in (\ref{nc}), and using the definition of Noether charge $Q_\xi$ (\ref{nch}), we obtain the
following relation
\be
\de H_\xi = \int_M J(\de ,\de_\xi ) = \int_{\partial M}\de Q_\xi -\xi\cdot\theta (\de )\, .\label{int2}
\ee
There are two contributions to the above expression, one at $S_\infty$ and the other
one at $S_\D$. As before, $\de E_\xi^{\infty}$, is the integral at the RHS of (\ref{int2}) calculated over
$S_\infty$, and  $\de E_\xi^{\D}$ the same integral calculated over $S_\D$.
Note that the necessary and sufficient condition for the existence of $H_\xi$ is the existence
of the form $B$ such that
\be
\int_{\partial M}\xi\cdot\theta (\de )=\de \int_{\partial M}\xi\cdot B\, .\label{condB}
\ee
Let us now consider how the different terms appearing in the action contribute to the Noether charges.

\subsubsection{Palatini action}

In this part let us consider the case of the Palatini action with boundary term. The symplectic potential current and 
the corresponding Noether charge form in this case are given by
\be
\theta_{\mathrm{PB}}(\de )=\frac{1}{2\k }\,\de\S^{IJ}\w\o_{IJ}\, ,\ \ \ \ Q_{\xi\mathrm{PB}} =\frac{1}{2\k}\, (\xi\cdot\S^{IJ})\w\o_{IJ}\, .
\ee
Due to the asymptotic behavior of the fields it follows that the contribution of the second term in (\ref{int2}) over $S_\infty$ vanishes. 
It follows that $B=0$ on $S_\infty$, which is consistent with the existence of  the ADM energy.
The remaining term at infinity in (\ref{int2}) is
\be
\de\int_{S_\infty}Q_{\xi\mathrm{PB}} =\frac{1}{2\k}\,\de\int_{S_\infty}(\xi\cdot\S^{IJ})\w\o_{IJ}\, ,
\label{adm}
\ee
and since $\int_{S_\infty}\de (\xi\cdot\S^{IJ})\w\o_{IJ}=0$, due to the asymptotic behaviour of the fields,
the above expression is equal to $\de E^\infty_\xi$ given in (\ref{energy_infinity}). Thus, in this case
\be
E_{\xi\rm ADM}=\int_{S_\infty}Q_{\xi\mathrm{PB}}\, ,
\ee
up to an additive constant that we choose to be zero. Note that a similar result is obtained in the second 
order formalism for the Einstein Hilbert action with the Gibbons-Hawking term, as
shown in \cite{pons}.

On the other hand, at the horizon the situation is different. In fact, it is easy to see that
\be
\int_{S_\D}Q_{\xi\mathrm{PB}} =0\, ,\quad {\textrm{and}}\quad  \int_{S_\D}\xi\cdot\theta_{\mathrm{PB}} (\de )
=\frac{1}{2\k}\,\int_{S_\D} (\xi\cdot\o_{IJ})\de\S^{IJ}\, .
\ee
Again, the necessary condition for the existence of $B$ such that (\ref{condB}) is satisfied, is that the surface gravity
$\k_{(\xi )}$ depends only on the area of the horizon \cite{afk}. 

We see that in this case
\be
\de E_\xi^{\infty}=\de\int_{S_\infty}Q_{\xi\mathrm{PB}}\quad ,\quad
\de E_\xi^{\D}=\int_{S_\D}\xi\cdot\theta_{\mathrm{PB}} (\de )\, .
\ee
We see then that the Hamiltonian energy and the Noether charge associated to the same vector field $\xi$, do not in general coincide \cite{crv1}.

\subsubsection{Holst and topological terms}

To end this section, let us present the Noether charges for the Holst term and the topological terms.
We will see that in most of these cases the integrals of the corresponding Noether charge 2-form
over $S_\infty$ and $S_\D$ vanish. The only exception is the Euler term that has a non-trivial contribution to the Noether charge.

Let us first consider the Holst term with its boundary term $S_{\mathrm{HB}}$,
given by (\ref{HolsttermBoundary}). We know that this term does not contribute to the Hamiltonian notion of energy, 
since it does not modify the symplectic structure of the Palatini action.
The Noether charge 2-form corresponding to $S_{\mathrm{HB}}$ is given by
\be
Q_{\xi\mathrm{HB}} = \f {1}{\k\g}(\xi\cdot\S^{IJ})\w\star\o_{IJ}=\f {1}{\co\g}(\xi\cdot e^I)\, \d e_I\, .
\ee
It can be checked that the Noether charges at infinity and the horizon vanish
\begin{equation}
\int_{S_\infty} Q_{\xi\mathrm{HB}} = \int_{S_\D} Q_{\xi\mathrm{HB}} = 0\, .
\end{equation}

On the other hand, we have seen in Section~\ref{sec:4.2} that we can also choose the Neih-Yan topological invariant
as a boundary term for the Holst term, resulting in the action $S_{\hny}$ that is cuadratic in torsion. 
The symplectic potential current for $S_{\hny}$ vanishes, as we saw in (\ref{HNY_symplectic_potential}), and
then from (\ref{noether}) it follows that the Noether charges vanish as well.

For the Pontryagin term (\ref{Pontryagin}) we obtain
\be
Q_{\xi\Po} =2 (\xi\cdot\o_{IJ})F^{IJ}\, .
\ee
In this case also the integral of the Noether charge 2-form $Q_{\xi\Po}$ over $S_\infty$ vanishes, 
\be
\int_{S_\infty} Q_{\xi\Po} =0\, ,
\ee
but on $S_\D$ we obtain 
\be\label{nch_po_hor}
\int_{S_\D} Q_{\xi\Po} =  
-2ic \int_{S_\D} (l\cdot V) \mathcal{R}\,\eps \, .
\ee
This expression is not gauge invariant on the horizon, under the rotations $m\to e^{i\theta}m$,  
the one-form $V$ transforms as $V\to V-i\d\theta$. So, in order to make the corresponding
Noether charge well defined we have to partially fix the gauge, by imposing $l\cdot\d\theta =0$.
This restricts the remaining gauge freedom 
$m\to e^{i\tilde{\theta}}m$ to the functions $\tilde{\theta}$ of the form  $\nabla_a\tilde{\theta}\,\heq\,w m_a+
\bar{w}\bm_a$, where $w$ is arbitrary.

On the other hand, we can calculate the Noether charge 2-form from the RHS of (\ref{Pontryagin}), and obtain 
\be
\tilde{Q}_{\xi\Po}=Q_{\xi\Po}-2\o^{IJ}\w\La_\xi\o_{IJ}\, .
\ee
It is easy to see that the integrals of the last term in the above equation over $S_{\infty}$ and $S_{\D}$ vanish,
due to our boundary conditions, hence the Noether charges remain invariant.

Similarly, for the Euler term we obtain
\be
Q_{\xi\Eu} =2(\xi\cdot\o_{IJ})\,{\star F}^{IJ}\, . 
\ee
Then, just as in the case of the Pontryagin term it is easy to see that 
\be
\int_{S_\infty}Q_{\xi\Eu} = 0\, ,
\ee
due to the asymptotic behaviour of the fields. 

At the horizon the situation is different since, due to the expressions (\ref{hor2}) and (\ref{dual_curvature_horizon}),
we obtain that the corresponding Noether charge is non vanishing
\be
\int_{S_\D}Q_{\xi\Eu} = 2c \int_{S_\D} (l\cdot\o )\, \mathcal{R}\,\eps =16\pi c\,\kappa_{(l)}\label{noether_euler_horizon}
\ee
since $l\cdot\o =\k_{(l)}$ is constant on the horizon and the remaining integral is a topological invariant of $S_\D$.
This result is consistent with the expression for the entropy of the Euler term in \cite{jacobson}, obtained in the second 
order formalism for stationary black holes. 
Though the Noether charge of the Euler term over a WIH is non-vanishing, the corresponding contribution to the energy is 
nonetheless, zero. As we have previously seen in Section~\ref{sec:6.1}, the variation of the energy at the horizon is 
\be
\de H_\xi^\D = \int_{S_\D}\de Q_\xi -\xi\cdot\theta (\de )\, ,
\ee
with $\xi =cl$. For the Euler term we obtain
\be 
\int_{S_\D} \xi\cdot\theta_\Eu (\de ) = 
16\pi c\,\de\kappa_{(l)}\, ,
\ee
and we see that this term cancels the variation of (\ref{noether_euler_horizon}) in  the expression for the energy at the horizon. 

Similarly as for the Pontryagin term, the variation of the RHS of (\ref{Euler}), leads to a change in the symplectic potential current and the Noether
charge 2-form, but the Noether charges stay invariant.

Let us end this section with a remark. As we have shown, the Noether charges at infinity of all the topological terms vanish for 
asymptotically flat boundary conditions,
but this is not the case for locally asymptotically anti-de Sitter (AAdS) space-times. In \cite{rodrigo},  
AAdS asymptotic conditions are considered and the Noether charge at 
infinity of the Palatini action with negative cosmological constant term turns out to be divergent.  
In that case the Euler term is added in order to make the action well defined
and finite. With this modification, the non vanishing (infinite) Noether charge becomes finite for
the well defined action.

\section{Discussion and remarks}
\label{sec:7}

In this contribution we have reviewed our understanding of first order gravity in the presence of boundaries.
Our focus was on action principles and the Hamiltonian covariant formalism that follows. When one is 
considering region with boundaries, or falloff conditions, it is important to explore the freedom that is 
available in the specification of the theory. In particular a rather well known `fact' is that `total derivatives 
do not matter' when constructing action principles. Since total derivatives can be converted into boundary 
terms, it is natural to explore terms of this type that can be added to the action, and their physical 
consequences. As general relativity is a prime example of a diffeomorphism invariant theory, the terms that one is 
allowed to add must satisfy this property (and not introduce new dynamical variables, either).
Thus, an important part of what we have reviewed are the well known topological terms that can be added to the action 
describing general relativity, together with other terms that are not topological but that do not change the equations of 
motion. As concrete examples of boundary conditions, we have considered asymptotically flat fall-off conditions at 
infinity, and/or isolated horizons boundary conditions at an internal boundary.
Our analysis was done using the covariant Hamiltonian formalism, that has
proved to be economical and powerful to unravel the Hamiltonian structure of
classical gauge field theories.
As a brief summary,  the main results that we have here presented can be put into four main categories.\\

i) We reviewed the covariant Hamiltonian formalism when boundaries are present. In particular, the treatment
that we recall here extends the standard formalism that appeared in the early literature that was tailored to the case without a boundary.\\

ii) We have reviewed the most general first order action for general
relativity in four dimensions. 
We described the additional boundary terms that one
needs to introduce to  have a differentiable action, which is
finite for the field configurations that satisfy our boundary conditions:
asymptotically flat spacetimes with an isolated horizon as an internal
boundary.\\

iii) We discussed the impact of the topological  and boundary terms added 
to have a well defined variational principle. In particular, we describe in detail their
contributions  to the symplectic
structure and the conserved Hamiltonian and Noether charges of the theory. We
showed that the topological terms do not modify the
symplectic structure. In the case of the Holst term (that is {\it not} topological), there is a particular 
instance in which it could modify the symplectic structure, and we discussed in detail the possible consequences of that choice. 
We have also shown that, for our boundary conditions,  the contribution from the Holst term to the 
Hamiltonian charges is always trivial.  Thus, the Hamiltonian structure of the theory remains
unaffected by the introduction of boundary and topological terms. However elementary this result may be, it
proves incorrect several assertions that have repeatedly appeared in the literature.\\

iv)  We have explored the relation between Hamiltonian and Noether charges. We review the result that shows
that these quantities do not in general coincide. Furthermore,
 even when the Hamiltonian conserved charges remain
insensitive to the addition of boundary and topological terms, the
corresponding Noetherian charges {\it do} depend on such choices. This has as
a consequence that the identification of Noether charges with, say, energy
depends on the details of the boundary terms one has added. For instance, if
one only had an internal boundary (and no asymptotic region), several
possibilities for the action are consistent, and the relation between energy
and Noether charge depends on such choices. We have also seen that the Pontryagin and Euler terms, contribute 
non-trivially to the Noether charge at the horizon.\\

\section*{Acknowledgements}

We would like to thank N. Bodendorfer, S. Deser,  T. Jacobson, R. Olea and J. D. Reyes for comments. 
This work was in part supported by CONACyT 0177840, DGAPA-UNAM IN100212, 
and NSF  PHY-1403943 and PHY-1205388 grants, the Eberly Research Funds of Penn State
and by CIC, UMSNH.

\end{document}